\documentclass[english,12pt,a4paper]{article}

\usepackage{babel}
\usepackage{amsmath}
\usepackage{amsfonts}
\usepackage{amssymb}
\usepackage{amsthm}
\usepackage{graphicx}

\usepackage{tikz}
\usetikzlibrary{calc,decorations.markings}

\usepackage{subfig}
\usepackage{placeins}

\usepackage[toc,page]{appendix}

\def\CC{\mathbb C} \def\RR{\mathbb R}   
 
\def\id{ 1\kern-4pt 1}

    \def\F{\mathcal F}

\def\pp(#1,#2){\frac{\partial #1}{\partial #2}}

  \def\c2{c^2}

\def\omithrj(**#1**){\null }
\def\nn{\nonumber}

\newcommand\nl{\hfill\break}
\newcommand\nll{\hfill\break\break}

\newtheoremstyle{break}
  {\topsep}{\topsep}%
  {\upshape}{}%
  {\bfseries}{}%
  {\newline}{}%

\theoremstyle{break}

  \def\ket{\rangle}
\def\ketindex#1{\ket \kern-2pt{\raisebox{-4pt}{$_#1$}}}
\def\roundbra{( \kern-3.5pt(}      \def\roundket{ ) \kern-3.5pt)}

\def \buildrelum#1\over#2{\mathrel{\mathop{\kern0pt #2}
                     \limits_{#1}   } }
\def \widetil{\mathrel\mathchar"0367}
\def \widesim{~~~~\lower.40em\hbox{$\widetil$}\mskip-10mu\succ ~~~}
\def \goes(#1){ \buildrelum #1 \over \longrightarrow}
\def \aslim(#1) {\buildrelum #1 \over \widesim}

\numberwithin{equation}{section}

\begin{document}

\centerline{\large \bf Space-time propagation of photon pulses in dielectric media,}  
\centerline{\large \bf illustrations with beam splitters}
\nl
\centerline{M. Federico, V. Dorier, S. Gu\'erin, H.R. Jauslin$^*$}
\nl
\centerline{Laboratoire Interdisciplinaire Carnot de Bourgogne - ICB, UMR 6303 CNRS}\\
\centerline{Universit\'e  Bourgogne Franche-Comt\'e, BP 47870, 21078 Dijon, France}
 \centerline{$^*$ e-mail: jauslin@u-bourgogne.fr, \qquad  \today}
\vskip20pt
{\it Published in: J. Phys. B 55 (2022) 174002} {\it  https://doi.org/10.1088/1361-6455/ac7e0e } 
\nll
\centerline{\bf  Coherent Control: Photons, Atoms and Molecules }  \centerline{\bf - a special issue honouring the life and work of Bruce W. Shore -}
%
\tableofcontents
\vskip2ex

\newpage

\begin{abstract}
Photons are the elementary quantum excitations of the electromagnetic field. Quantization is  usually constructed on the basis of an expansion in eigenmodes, in the form of plane waves. Since they form a basis, other electromagnetic configurations can be constructed by linear combinations. In this presentation we discuss a formulation constructed in the general formalism of bosonic Fock space, in which the quantum excitation can be constructed directly on localized pulses of arbitrary shape. 
Although the two formulations are essentially equivalent, the direct formulation in terms of pulses has some conceptual and practical advantages, which we illustrate with some examples. The first one is   the passage of a single photon pulse through a beam splitter. The analysis of this formulation in terms of pulses in Fock space shows that there is no need to introduce ``vacuum fluctuations entering through the unused port'', as is often done in the literature. Another example is the Hong-Ou-Mandel effect. It is described as a time dependent process in the Schr\"odinger representation in Fock space. The analysis shows explicitly how  the two essential ingredients of the Hong-Ou-Mandel effect are the same shape of the pulses and the bosonic nature of photons. 
This formulation shows that all the phenomena involving linear quantum optical devices  can be described and calculated on the basis of the time dependent solution of the corresponding classical Maxwell's equations for pulses, from which the quantum dynamics in Fock space can be immediately constructed.
\end{abstract}

\section{Introduction}

Recent progress in the techniques for the production and in the detection of single photons has allowed to create photons at demand and heralded with pulse shapes  that can be engineered to a large extent 
\cite{Kuhn2002,McKeever2004,Wilk2007,Ritter2012,Mucke2013,Boozer2007,Kuhn2010,Dilley2012,Kuhn2015}.
These photon pulses can be made to propagate through different optical devices \cite{Saharyan2021}
\cite{Leonhardt2003}, like semi-transparent mirrors, beam splitters, lenses, and optical fibers,  and then detected with space-time resolution.
In this article we describe a framework for the propagation of single- or few-photon pulses through linear optical media  for which dissipation and dispersion can be neglected in the range of considered frequencies. The presented approach is based on the pioneering work of Glauber and Lewenstein \cite{Glauber1991} \cite{Wubs2003}, combined with Fock space tools of quantum field theory
\cite{Berezin1966} \cite{Debievre2006} \cite{Debievre2007}. A key role is played by the {\it frequency operator}, which contains the main information for the dynamics, and makes the link between the classical Maxwell equations and the quantum theory. 
{The interaction of propagating photon pulses with matter has been addressed in the literature from different perspectives, as described e.g. in \cite{Fabre2008} \cite{Fabre2020} \cite{Kiilerich2019}\cite{Kiilerich2020}
and in the references therein.}

\section{Passive inhomogeneous  dielectric  media}

A passive dielectric described by a dielectric coefficient $\epsilon(\vec x) > 0$, is a linear medium for which the dissipation and the dispersion can be neglected in the considered frequency range. The corresponding wave equation, which is equivalent to the macroscopic phenomenological Maxwell equations, can be written as 
\begin{eqnarray}
\epsilon(\vec x)
\frac{\partial^2\vec A}{\partial t^2}
&=&  -c^2 \nabla\times\nabla\times \vec A \\
\phantom{xxxxx}\vec E &=& - \frac{\partial \vec A}{\partial t}, \qquad
\vec B =  \nabla\times \vec A,
\end{eqnarray}
where $\vec A$ is the vector potential in the generalized Coulomb gauge \cite{Glauber1991} \cite{Wubs2003}, which satisfies the generalized transversality constraint $\nabla\cdot(\epsilon \vec A)=0$.
In order to put this equation in Hamiltonian form with a self-adjoint operator
\cite{Glauber1991} \cite{Wubs2003}, we make the following change of variables:
\begin{equation}
\vec A':=\sqrt{\epsilon} \vec A,
\end{equation}
which leads to 
\begin{equation}
\label{dielectric_wave_eq333}
\frac{\partial^2\vec A'}{\partial t^2} =-  \frac{1}{\sqrt{\epsilon}}  c^2 \nabla\times\nabla\times \frac{1}{\sqrt{\epsilon}}~\vec A'.
\end{equation}
The operator 
\begin{equation}
\frac{{c}}{\sqrt{\epsilon}} \nabla\times\nabla\times  \frac{{c}}{\sqrt{\epsilon}} =
\Xi^\dag \Xi, \qquad {\rm with~} \Xi= \nabla\times\frac{{c}}{\sqrt{\epsilon}}
\end{equation}
is positive and selfadjoint, and thus  there is a unique positive operator $\Omega$, such that 
\begin{equation} \label{def-Omega}
\Xi^\dag \Xi=\Omega^2, \qquad {\rm i.e.~} \Omega= \left(\frac{{c}}{\sqrt{\epsilon}} \nabla\times\nabla\times  \frac{{c}}{\sqrt{\epsilon}}\right)^{1/2}.
\end{equation} 

Thus Maxwell's equations in a passive dielectric medium 
can be written as
\begin{equation}
\label{dielectric_wave_eq33}
\frac{\partial^2\vec A'}{\partial t^2} =-\Omega^2 \vec A'.
\end{equation}
which has the same structure as the ones in empty space, where $\Omega_0^2=-c^2\Delta$, but with the {\it frequency operator}  $\Omega$ defined in \eqref{def-Omega}  and the generalized transversality constraint. Thus we can formulate the Hamiltonian structure for the passive dielectric medium as follows.

\subsection{Hamiltonian structure of  Maxwell's equations  for a passive    inhomogeneous  dielectric medium}
By defining 
\begin{equation}
\vec \Pi' := \epsilon_0 \frac{\partial \vec A'}{\partial t}  
\end{equation}
as the  momentum canonically conjugate to $\vec A'$, the wave equation can be written in Hamiltonian form
\begin{eqnarray}
 \frac{\partial\vec A'}{\partial t}
 & = & \frac{\delta H}{\delta \vec \Pi'(\vec x)} =
 \frac{1}{\epsilon_0} \vec \Pi'  \\  
  \frac{\partial\vec \Pi'}{\partial t}
  & = &  {-}\frac{\delta H}{\delta \vec A'(\vec x)} =  -\epsilon_0\Omega^2  \vec A'.
\end{eqnarray}
with the Hamilton functional   
\begin{eqnarray} \label{cl-Hamiltonian}
H&=&\int_{\RR^3}d^3x~\biggl( \frac{1}{2\epsilon_0}\vec \Pi'(\vec x)\cdot \vec \Pi'(\vec x) + \frac{\epsilon_0}{2} \vec A'(\vec x)\cdot \Omega^2  \vec A'(\vec x)\biggr).
\end{eqnarray}
and the constraints
\begin{eqnarray} \label{gen-constraints}
 \nabla\cdot \left( \sqrt{\epsilon}\vec A' \right) =0, \quad \nabla\cdot  \left( \sqrt{\epsilon}\vec \Pi' \right) =0.
 \end{eqnarray}

\subsection{Quantization of the electromagnetic field 
 in passive    inhomogeneous  dielectric media}
The electromagnetic field in a passive inhomogeneous  dielectric medium can be quantized in the same way
as in free space \cite{Glauber1991} 
\cite{Wubs2003} 
\cite{Berezin1966}
\cite{Debievre2006}
\cite{Debievre2007}.
The only difference is the form of the operator $\Omega^2$, and the generalized transversality constraints 
\eqref{gen-constraints}.
Our presentation is based on the general formulation of \cite{Debievre2006} \cite{Debievre2007}, adapted to the case of the electromagnetic field, which requires to take into account the vector character and the transversality constraints. This formulation is particularly well-adapted for the space-time description of the propagation of the photon states. The main steps can be summarized as follows.
\nll
A) By introducing the following complex representation of the fields 
 \begin{eqnarray}
\label{ClassicalWaveComplex}
\vec \psi:=\frac{1}{\sqrt{2\hbar}} \biggl((\epsilon_0\Omega)^{1/2}\vec A '+ i (\epsilon_0\Omega)^{-1/2}\vec \Pi' \biggr),
\end{eqnarray}
the classical Maxwell equations can be written equivalently as
 \begin{eqnarray} \label{ClassicalWaveEq}
i \frac{\partial \vec \psi}{\partial t} =\Omega   \vec \psi.
\end{eqnarray}
B) The classical phase space $\vec\Pi',\vec A'$ has a natural Hilbert space structure, which in the complex representation takes the form
\begin{eqnarray}
\mathcal{H}_{cl} := \{ \vec \psi(\vec x)  ~|~ \langle \vec \psi|\vec\psi \rangle <\infty\},
\end{eqnarray}
with the scalar product
\begin{eqnarray} \label{def-scalar-product}
 \langle \vec \psi_A | \vec \psi_B \rangle := \int_{\RR^3}d^3 x~ \vec \psi_A^*(\vec x) \cdot \vec \psi_B (\vec x).
\end{eqnarray}
The fields in this space are essentially fields with a finite electromagnetic energy.
\nll
{\bf Nomenclature:} We will use the following nomenclature, based on the one of Refs. 
\cite{Fabre2008}\cite{Fabre2020}: A {\it classical mode} is a configuration of the classical electromagnetic field described equivalently by $[\vec E(\vec x),\vec A(\vec x)]$, $[\vec E(\vec x),\vec B(\vec x)]$ or the complex representation $\vec \psi(\vec x)$. 
The classical modes can be normalized using the scalar product \eqref{def-scalar-product}. A classical mode, also called {\it non-monochromatic mode} can be viewed as an equivalence class (i.e. up to normalization) of possible initial conditions for Maxwell's equations.  {\it Classical eigenmodes} $\vec\varphi_\kappa$ , called also  {\it classical normal modes},
are a particular type of modes that are eigenfunctions of the frequency operator, i.e. satisfying
\begin{eqnarray}
\Omega^2 \vec\varphi_\kappa = \omega_\kappa^2 \vec\varphi_\kappa.
\end{eqnarray}
C) Based on this classical Hilbert space one can construct the associated bosonic Fock space ${\F}^{\mathfrak{B}}(\mathcal{H}_{cl})$, defined as follows \cite{Berezin1966}
\cite{Debievre2006}
\cite{Debievre2007}
\cite[Sect. 18.5]{Honegger2015}:
\begin{eqnarray}
{\F}^{\mathfrak{B}}(\mathcal{H}_{cl}) & := & 
\oplus_{m=0}^{\infty}{\F}^{\mathfrak{B}}_m, \\
\nonumber \\
{\F}^{\mathfrak{B}}_0& := & \CC \\
{\F}^{\mathfrak{B}}_1 & := &  \mathcal{H}_{cl} \\
{\F}^{\mathfrak{B}}_2 & := &  \hat P_2 \left(  \mathcal{H}_{cl} \otimes \mathcal{H}_{cl} \right) 
\equiv  \mathcal{H}_{cl} \otimes_S \mathcal{H}_{cl} \\
&\vdots & \nonumber\\
{\F}^{\mathfrak{B}}_{m} & := &  \hat P_m \left( \mathcal{H}_{cl} \otimes \mathcal{H}_{cl}  \otimes \ldots \otimes\mathcal{H}_{cl} \otimes \mathcal{H}_{cl}  \right) \\
&\vdots & \nonumber
\end{eqnarray}
where $\hat P_m$ are the symmetrization operators,  for instance $\hat P_2 |\vec\psi_1\otimes \vec\psi_2\rangle \equiv |\vec\psi_1\otimes_S \vec\psi_2\rangle := \frac{1}{2}  (|\vec\psi_1\otimes \vec\psi_2 \rangle+ |\vec\psi_2\otimes \vec\psi_1\rangle) $. 
\\ \\
The vectors in this Fock space can be represented as $(\phi_0,\phi_1, \phi_2,\ldots )$, where $\phi_m\in {\F}^{\mathfrak{B}}_{m}$ is the $m-$photon component.
The normalized {\it vacuum state} is defined up to a phase factor as
$
|\varnothing \rangle := (1,0, 0,\ldots ).              
$
\nll
On this symmetric Fock space one can construct, for each  $\vec\psi\in\mathcal{H}_{cl}$, the following creation-annihilation operators
\begin{equation}
{\hat B}_{\vec\psi} : \F^{\mathfrak{B}}_m \to \F^{\mathfrak{B}}_{m-1} \qquad {\hat B}^\dag_{\vec\psi} : \F^{\mathfrak{B}}_m \to \F^{\mathfrak{B}}_{m+1}
\end{equation}
 defined by their action 
\begin{eqnarray}
\label{415-creation-op} 
\kern-50pt {\hat B}^\dag_{\vec\psi} | \vec\psi_1\otimes\ldots\otimes\vec\psi_m \rangle & := &\sqrt{m+1}  ~\hat P_{m+1}| \vec\psi\otimes \vec\psi_1\otimes\ldots\otimes\vec\psi_m \rangle \\
\label{415-annihilation-op}
 \kern-50pt  {\hat B}_{\vec\psi} | \vec\psi_1\otimes\ldots\otimes\vec\psi_m \rangle & := & ~\frac{1}{\sqrt{m} }\sum_{j=1}^m \langle\vec\psi | \vec\psi_j\rangle~\hat P_{m-1}~ |  \vec\psi_1\otimes\ldots\otimes\check{\vec{\psi}}_j\ldots\otimes\vec\psi_m \rangle,
 \end{eqnarray}
 where the notation $\check{\vec{\psi}}_j$ indicates that this term is missing.
 \nll
  {\bf Examples of one- and two-photon states:} If $\vec \psi(\vec x)$ is a pulse-shaped field, then 
 \begin{eqnarray} \label{one-photon-psi-x}
 {\hat B}^\dag_{\vec\psi}|\varnothing\rangle = | \vec\psi\rangle = \vec\psi(\vec x)
 \end{eqnarray}
  is a one-photon state carried by the classical mode  $\vec \psi(\vec x)$, described in the position space representation. 
 \begin{eqnarray}
 {\hat B}^\dag_{\vec\psi_A} {\hat B}^\dag_{\vec\psi_B} |\varnothing\rangle =\sqrt{2}~ | \vec\psi_A \otimes_S \vec\psi_B \rangle =  \frac{1}{\sqrt{2}}  (|\vec\psi_A\otimes \vec\psi_B\rangle+ |\vec\psi_B\otimes \vec\psi_A\rangle)
 \end{eqnarray}
  is a two-photon state carried by the classical modes $\vec\psi_A$ and $\vec\psi_B $.
  \\
  The annihilation operators act on 0, 1 and 2-photon states by
 \begin{eqnarray}
 {\hat B}_{\vec\psi}|\varnothing\rangle &= 0 \\
  {\hat B}_{\vec\psi_A} | \vec\psi_B\rangle  &= \langle \vec\psi_A | \vec\psi_B  \rangle~ |\varnothing\rangle \\
  {\hat B}_{\vec\psi_A} | \vec\psi_B \otimes_S \vec\psi_C \rangle  &=  \frac{1}{\sqrt{2}} 
  \left(  \langle \vec\psi_A| \vec\psi_B \rangle    ~| \vec\psi_C\rangle 
  + \langle \vec\psi_A | \vec\psi_C\rangle ~| \vec\psi_B\rangle  \right).
 \end{eqnarray}
  The bosonic creation-annihilation operators satisfy the following commutation relations:
  \begin{eqnarray} \label{commBBdag}
\left[  {\hat B}_{\vec\psi_A}, {\hat B}^\dag_{\vec\psi_B} \right]  & = & \langle {\vec\psi_A} | {\vec\psi_B} \rangle  \\
\left[   {\hat B}_{\vec\psi_A}, {\hat B}_{\vec\psi_B} \right]   = & 0 & =
\left[   {\hat B}^\dag_{\vec\psi_A}, {\hat B}^\dag_{\vec\psi_B} \right]. 
\end{eqnarray}
We remark that in the particular case where $\vec\varphi_A$ and $\vec\varphi_B$ are normalized vectors orthogonal to each other, for instance if $\vec\varphi_A$ and $\vec\varphi_B$ are  orthonormal eigenmodes of a cavity, the commutation relation \eqref{commBBdag} takes the form
  \begin{eqnarray} 
\left[  {\hat B}_{\vec\varphi_A}, {\hat B}^\dag_{\vec\varphi_B} \right]  & = & \delta_{A,B},
\end{eqnarray}
which is the  commutation relation that is traditionally used in the treatment of systems of harmonic oscillators and in cavity quantum electrodynamics. The commutation relations \eqref{commBBdag} are a slight generalization \cite[p.105, Eq.(3.192)]{Chiao2014}, \cite[Eq. (38)]{Fabre2020}, in which the classical carrying modes of the photons are not necessarily normal modes nor orthonormal. As we will see in the applications to beam splitters this generalization is very useful for the description of the propagation of pulse-shaped photon states.
 \nll
 
{  {\bf Remark:}
In Eqs.  (2.24), (2.25)  \eqref{415-creation-op}   \eqref{415-annihilation-op} the creation-annihilation operators $\hat B^\dag_{\vec\psi}, \hat B_{\vec\psi}$ are labeled by 
fields $\vec\psi$ that belong to the single-photon Hilbert space $\mathcal{H}_{cl} $, i.e. which are square integrable and have a finite classical energy. The definition can be extended formally to include the continuum eigenfunctions 
$\vec\varphi_\kappa$ of the frequency operator $\Omega$, that are not in the Hilbert space,    (e.g. plane waves in the case without a medium).
When applied to the vacuum, $\hat B^\dag_{\vec\varphi_\kappa} | \varnothing\rangle = |\vec\varphi_\kappa\rangle$,
can be interpreted formally as a single-photon state carried by a monochromatic plane wave. These objects are useful e.g. for the mathematical  representation of the electromagnetic field observables \eqref{Ac_efct}\eqref{Pic_efct}\eqref{energy-observable}, since the plane waves are a ``generalized basis'' of the Hilbert space. They are also useful to represent the pulse states as linear combinations of plane waves,
 \begin{eqnarray}
 \kern-50pt 
|\psi\rangle = \hat B^\dag_{\vec\psi} | \varnothing\rangle 
=\hat B^\dag_{\int d\kappa~z(\kappa)\vec\varphi_\kappa} | \varnothing\rangle
=\int d\kappa~z(\kappa)\hat B^\dag_{\vec\varphi_\kappa} | \varnothing\rangle
=\int d\kappa~z(\kappa) |\vec\varphi_\kappa \rangle,
 \end{eqnarray}
 where we have used the linearity property \eqref{B-linearity} of the creation operators.
However, $ |\vec\varphi_\kappa\rangle$ is not a single-photon state that can be created in the laboratory, since it would contain an infinite energy, and it would occupy uniformly the whole physical space $\RR^3$. Thus, purely monochromatic fields, although mathematically useful, cannot be physical single-photon states. This is reflected in the theory by the fact that they do not belong to the Hilbert space.  In experiments what is often created are almost-monochromatic photon states, that can be represented by a monochromatic plane wave multiplied by a pulse-shaped envelope, which is what we use below for the illustrations with beam splitters. They have a finite energy and they  belong to the Hilbert space of square integrable fields $\mathcal{H}_{cl} $.  We remark that this issue is different in free space than in a cavity, where purely monochromatic single-photon states can in principle be created, since the volume and thus the energy are finite.  }
\nll
 D) The basic electromagnetic field observables are represented by the following operators acting on the bosonic Fock space. We start with a complete orthonormal set of continuum eigenfunctions $\vec\varphi_\kappa$ of the frequency operator,
 \begin{eqnarray}
 \Omega^2 \vec\varphi_\kappa = \omega^2_\kappa \vec\varphi_\kappa, \qquad \omega_\kappa >0.
 \end{eqnarray}
 The electromagnetic field operators can be expressed as
 \begin{eqnarray}
\label{Ac_efct}   
\sqrt{\epsilon} \vec {\hat A}=\vec {\hat A'}(\vec x)   & = &\sqrt{\frac{\hbar}{2\epsilon_0}}  \int d\kappa~  
  \omega_\kappa^{-1/2} \bigl(  \vec\varphi_\kappa(\vec x)~\hat{B}_{\vec\varphi_\kappa}+ \vec\varphi^{\kern0.3ex *}_\kappa(\vec x)~\hat{B}_{\vec\varphi_\kappa}^\dag \bigr) \\
  \label{Pic_efct}
-\epsilon_0\sqrt{\epsilon}\vec{\hat  E}=\vec {{\hat {\Pi'}}}(\vec x) & = & -i\sqrt{\frac{\hbar\epsilon_0}{2}} \int d\kappa~
\omega^{1/2}_\kappa \bigl(  \vec\varphi_\kappa(\vec x)~\hat{B}_{\vec\varphi_\kappa} - \vec\varphi^{\kern0.3ex *}_\kappa(\vec x)~\hat{B}_{\vec\varphi_\kappa}^\dag \bigr).
\end{eqnarray}
and the total energy operator (after normal ordering) as
\begin{eqnarray}\label{energy-observable}
\hat H & = & \int d\kappa~ \hbar \omega_\kappa \hat B_{\vec\varphi_\kappa}^\dag  \hat B_{\vec\varphi_\kappa}. 
\end{eqnarray}
The notation $ \int d\kappa$ is to be interpreted as an abbreviation of a combination of integrals and discrete sums on the labels of the eigenfunctions. For instance in the case of the free field $\vec\varphi_\kappa(\vec x) =(2\pi)^{-3/2} \vec \varepsilon_s(\vec k) e^{i\vec k\cdot \vec x}$,  $\kappa=(\vec k, s)$, where $\vec k\in\RR^3$ is the wave vector, $s=\pm$ is an index for the two polarizations, and $\int d\kappa~ \equiv  \int_{\RR^3}d^3k\sum_{s=\pm}$. The explicit link with the usual formulas of the observables for the free field is described in Appendix \ref{Appendix-reciprocal}.

\section{Evolution of the photon states determined   from the   dynamics of the classical modes   
\label{Section-evolution-photon-states} }

We are going to show that the time evolution of photon states in Fock space, \nl determined by the equation
\begin{eqnarray} \label{SchrodingerFockEq}
i\hbar\frac{\partial |\Phi \rangle}{\partial t}& = & \hat H |\Phi\rangle, \qquad |\Phi \rangle\in \mathcal{F}^{\mathfrak{B}}
\end{eqnarray}
 can  be expressed in terms of the time evolution of the classical modes, determined by the classical wave equation, i.e. by Maxwell's equations. 
 \\
 {\bf  (a)} In particular, for a one-photon initial condition $| \Phi(t=0)\rangle= \hat B_{\vec\psi(t=0)}^\dag|\varnothing\rangle $
the time evolution is
\begin{eqnarray}  \label{onepev}
| \Phi(t) \rangle& = & \hat B_{\vec\psi(t)}^\dag|\varnothing\rangle,
\end{eqnarray}
where $\vec\psi(t)$ is the solution of the classical equation \eqref{ClassicalWaveEq}  with initial condition $\vec\psi(t=0)$.  
\nll
{\bf  (b)} For a two-photon initial condition
 $| \Phi(t=0)\rangle= \hat B_{\vec\psi_B(t=0)}^\dag \hat B_{\vec\psi_A(t=0)}^\dag|\varnothing\rangle $ the time evolution is 
\begin{eqnarray} \label{twopev}
 | \Phi(t)\rangle= \hat B_{\vec\psi_B(t)}^\dag \hat B_{\vec\psi_A(t)}^\dag|\varnothing\rangle, 
\end{eqnarray} 
where $\vec\psi_A(t)$ and  $\vec\psi_B(t)$ are the solutions of the classical equation \eqref{ClassicalWaveEq} for the corresponding initial conditions.
\nll
{\bf  (c)} In the general case of an N-photon initial condition
 $| \Phi(t=0)\rangle= \prod_{j=1}^N \hat B_{\vec\psi_j(t=0)}^\dag |\varnothing\rangle $ the time evolution is 
\begin{eqnarray} \label{Npev}
 | \Phi(t)\rangle= \prod_{j=1}^N \hat B_{\vec\psi_j(t)}^\dag |\varnothing\rangle, 
\end{eqnarray} 
where $\vec\psi_j(t)$  are the solutions of the classical equation \eqref{ClassicalWaveEq} for the corresponding initial conditions.
\nll 
A proof of the time evolution formulas \eqref{onepev},\eqref{twopev} and \eqref{Npev} is given in Appendix  \ref{Appendix-proof-evolution-photon-states}.
\nll
{\bf Remark:}   The family of operators $\hat B_{\vec\psi(t)}^\dag$ parametrized by time can be thought of as a time-dependent operator
which can be written as
 \begin{eqnarray} \label{notHeisenberg}
\hat B_{\vec\psi(t)}^\dag= \hat U(t)~\hat B_{\vec\psi(t=0)}^\dag ~\hat U^\dag(t)
\end{eqnarray} 
where $ \hat U(t) = e^{-\frac{i}{\hbar} \hat H~t}$ is the propagator in Fock space. This can be verified e.g. by applying both sides to the vacuum state: Since
 \begin{eqnarray} 
\hat U^\dag(t) | \varnothing  \rangle  &=& \sum_{n=0}^\infty \left( \frac{i}{\hbar} t \right)^n  \left( \sum_{\kappa} \hbar \omega_\kappa 
\hat B_{\vec\varphi_{\kappa}}^\dag \hat B_{\vec\varphi_{\kappa}} \right)^n  | \varnothing  \rangle =  | \varnothing  \rangle 
\end{eqnarray} 
we can write
 \begin{eqnarray} 
  \hat U(t)~\hat B_{\vec\psi(t=0)}^\dag \hat U^\dag(t)   | \varnothing  \rangle &=&
 \hat U(t)~\hat B_{\vec\psi(t=0)}^\dag  | \varnothing  \rangle  =  \hat U(t)~   | \vec\psi(t=0)  \rangle  \\
&=& | \vec\psi(t)  \rangle 
= \hat B_{\vec\psi(t)}^\dag  | \varnothing  \rangle.
 \end{eqnarray} 
We remark however that it is not the time evolution in the Heisenberg picture, which is given by a different expression 
\cite[p.84]{Chiao2014}:
 \begin{eqnarray} \label{Heisenberg}
\left(\hat B_{\vec\psi(t=0)}^\dag  \right)^{{\rm Heisenberg}}(t)= \hat U^\dag(t)~\hat B_{\vec\psi(t=0)}^\dag ~\hat U(t) 
=\hat B_{\vec\psi(-t)}^\dag
\neq \hat B_{\vec\psi(t)}^\dag.
\end{eqnarray} 

\section{One-photon pulse propagation through a  beam splitter \label{Section-one-photon-beam splitter}}

We consider a single pulse   arriving on a 50-50\% beam splitter at $45^o$. 

\vskip 4ex
\begin{minipage}[t]{0.3\textwidth}
\begin{tikzpicture} [scale=3]
\draw [line width=7pt,gray] (-0.415,-0.4) --  (0.385,0.4);
\draw [->,blue](-0.03,0.03) --  (-0.03,1);
\draw [->,blue](-1,0.03) --  (-0.5,0.03); \draw [->,blue](-0.5,0.03) --  (-0.03,0.03);
\draw [->,blue](-0.03,0.03) --  (1,0.03);
\draw [line width=3pt]  (0.2,1.1) -- (-0.2,1.1)   arc(180:0:0.195) ;
\draw [line width=3pt]  (1.1,0.2) -- (1.1,-0.2)   arc(-90:90:0.195) ;
\draw [domain=-0.8:-0.2,blue] plot (\x,{0.1+0.3*sin((\x+0.8)*pi/(-0.2+0.8) r)^4});
\node at (-0.5,0.5) {$ {\color{blue} \vec\psi_{i} }$ };
\node at (0.1,1.0) {$y$};  \node at (1.0,0.1) {$x$};
\node at (0.35,1.2) {$D_Y$};  \node at (1.42,0.2) {$D_X$};
\node at (-0.8,1.2) {(a)}; 
\end{tikzpicture}
\end{minipage}
\hskip15ex
\begin{minipage}[t]{0.55\textwidth}
\begin{tikzpicture} [scale=3]
\draw [line width=7pt,gray] (-0.415,-0.4) --  (0.385,0.4);
\draw [->,blue](-0.03,0.03) --  (-0.03,1);
\draw [->,blue](-0.8,0.03) --  (-0.5,0.03); \draw [->,blue](-0.5,0.03) --  (-0.03,0.03);
\draw [->,blue](-0.03,0.03) --  (1,0.03);
\draw [line width=3pt]  (0.2,1.1) -- (-0.2,1.1)   arc(180:0:0.195) ;
\draw [line width=3pt]  (1.1,0.2) -- (1.1,-0.2)   arc(-90:90:0.195) ;
\draw [domain=0.2:0.8,blue] plot (\x,{0.1+0.15*sin((\x-0.2)*pi/(0.8-0.2) r)^4});
\node at (0.8,0.3) {$ {\color{blue} \vec\psi_{T} }$ };
\draw [domain=0.2:0.8,blue] plot ({-0.1-0.15*sin((\x-0.2)*pi/(0.8-0.2) r)^4}, \x);
\node at (-0.3,0.8) {$ {\color{blue} \vec\psi_{R} }$ };
\node at (0.1,1.0) {$y$};  \node at (1.0,0.1) {$x$};
\node at (0.35,1.2) {$D_Y$};  \node at (1.42,0.2) {$D_X$};
\node at (-0.7,1.2) {(b)}; 
\end{tikzpicture}
\end{minipage}
\vskip 1ex
\noindent
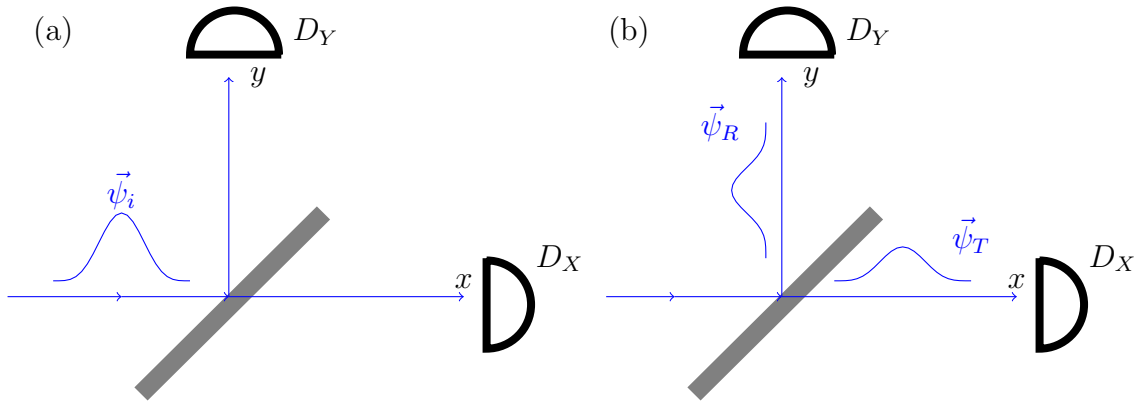
Figure 1. Schematic illustration of the partial transmission and reflection of a single-photon state through a 50\%-50\% beam splitter. (a) Incoming pulse, (b) reflected and transmitted pulses. The dynamics of the single-photon state is the same as the classical dynamics of pulses through beam splitters according to Eq. \eqref{onepev}. $D_X$ and $D_Y$ represent detectors.
\vskip 4ex

\subsection{Description of the initial classical mode and its time evolution}

The pulse at the initial time $t_i$, before the pulse arrives at the beam splitter, is a classical mode.  \\
We define the following {\it pulse shape function} $\vec S( r_1,r_2,r_3,t;~ k) $, depending on three spatial arguments, $r_1,r_2,r_3$, a temporal variable $t$,  and  a wave number $k$,
\begin{eqnarray}   \label{pulse-shape}
\vec S( r_1,r_2,r_3, k, t) & := &\mathcal{N} \vec e_z \mathcal{E}(r_1-ct) e^{i(k r_1-\omega t)} g(r_2,r_3)
\end{eqnarray}
where $\vec e_z=(0,0,1)$ is the linear polarization vector,  $\omega =ck>0$  is the carrier frequency, and
$ \mathcal{E}(r_1)$ is the pulse envelope in the direction of propagation.  $g(r_2,r_3)$ is the transverse profile, which we assume to be smaller than the beam splitter, and in the usual circumstances does not change significantly with the propagation. 
 For the pulse   envelope $\mathcal{E}(r_1)$ one can take e.g.  a $\sin^2$ with a finite support, so that there is no ambiguity  about when the process starts, or a Gaussian multiplied by a characteristic function to give it a finite support.
$\mathcal{N}$ is a normalization constant, chosen such that $\int_{\RR^3} d^3 r ~ \vec S^*\cdot \vec S= 1$.
\\
With these assumptions the propagation of the photon pulse can be reduced to a one dimensional propagation.
\nll
The initial mode is chosen as
\begin{eqnarray}  \label{SA1}
\vec\psi_{i}(\vec x) &=  &  \vec S(x,y,z;~ k, t=0) =:  \vec S(x,t=0),
\end{eqnarray}
where we introduce an abridged notation $\vec S(x,t)$, indicating only the first spatial argument of its direction of propagation and the time argument. 
\nll
After crossing  the beam splitter, the classical mode evolves into  a reflected and a transmitted pulse (see Figure 1)
\begin{eqnarray}
& & \vec\psi_{i} \rightarrow\vec\psi_{R}  +  \vec\psi_{T}  \\
\vec\psi_{R}(\vec x,t) & = & \tilde  r~  \vec S(y-ct,t) =: \tilde  r~   \vec Y(y) \\
 \vec\psi_{T}(\vec x,t) & = &  \tilde t~   \vec S(x-ct,t) =: \tilde t~ \vec X(x), 
\end{eqnarray}
where  $ \tilde r, \tilde t$ are the reflection and transmission coefficients, which in general satisfy  the relations  $ |\tilde r|^2+  |\tilde t|^2=1$ and 
 $ \tilde r^*  \tilde t+ \tilde r  \tilde t^*=0$.  
We have introduced the abridged notations $ \vec Y (y), \vec X(x)$  to improve the readability  of the construction below. ($ \vec X$: propagation along the $x$-axis; $ \vec Y$: propagation along the $y$-axis).

\subsection{Quantum dynamics of the one-photon state }

We assume that the system is prepared with one photon on the initial mode, described  in the Fock space by
\begin{eqnarray} \label{initstate1}
 | \Phi_i\rangle &=&\hat B^\dag_{\vec \psi_{i}}   |\varnothing\rangle  = 
  | \vec\psi_{i}   \rangle.
 \end{eqnarray}
The mode function must be normalized:
\begin{eqnarray} 
\int d^3 x   |\vec \psi_{i}(\vec x)|^2 =1,
\end{eqnarray}
which entails 
\begin{eqnarray} 
\left[\hat B_{\vec \psi_{i}}~,~ \hat B^\dag_{\vec \psi_{i}}  \right]&=&\id.
\end{eqnarray}
 This implies that the initial state \eqref{initstate1} is normalized, $\langle  \Phi_i  | \Phi_i\rangle=1$.
\nll
After crossing  the beam splitter, this one-photon initial state evolves  to 
\begin{eqnarray} 
 | \vec\psi_{i}   \rangle \rightarrow 
&&| \vec\psi_{R}  +  \vec\psi_{T} \rangle    \equiv     \hat B^\dag_{ \vec\psi_{R}  +  \vec\psi_{T} }  |\varnothing\rangle    \label{1photonevolution} \\ 
 &= &|  \tilde r ~  \vec Y(y)  +   \tilde t ~\vec X(x) \rangle.
 \end{eqnarray}
  The physical interpretation of the state at time $t$ after the crossing of the beam splitter is as follows: $| \vec\psi_{R}  +  \vec\psi_{T} \rangle  $
 is a one-photon state on the single classical mode $\vec\psi_{R}  +  \vec\psi_{T}$, which has two spatially disjoint components, one propagating in the $x$ direction and the other one in the $y$ direction (see Figure 1).
 
\subsection{Remarks on the  literature on the beam splitter, involving a    fourth port representing   ``incoming fluctuations of the \mbox{vacuum}" }
 
 First we remark that the creation operator in \eqref{1photonevolution} can be decomposed as the sum of two terms
 \begin{eqnarray} \label{separateB}
    \hat B^\dag_{ \vec\psi_{R}  +  \vec\psi_{T} }  =   \hat B^\dag_{ \vec\psi_{R}} +  \hat B^\dag_{ \vec\psi_{T} }. 
 \end{eqnarray}
 This property is a direct consequence of the definition \eqref{415-creation-op}, due to the linearity of the tensor product in each of its arguments.
Since the time evolution of the classical modes is unitary,  it implies that the norm of the states in Fock space is preserved, and also 
 \begin{eqnarray} \label{globalCR}
\left[\hat B_{ \vec\psi_{R}  +  \vec\psi_{T}}~,~ \hat B^\dag_{ \vec\psi_{R}  +  \vec\psi_{T}}  \right]&=&\id.
\end{eqnarray}
However, the separate terms in \eqref{separateB} satisfy the commutation relations
\begin{eqnarray}  \nn
\left[\hat B_{\vec\psi_{R}}~,~ \hat B^\dag_{\vec\psi_{R}}  \right]&=& \langle\vec\psi_{R}~|~ \vec\psi_{R} \rangle=  |\tilde r|^2  \neq \id  \\ \label{separateCR}
\left[\hat B_{\vec\psi_{T}}~,~ \hat B^\dag_{\vec\psi_{T}}  \right]&=& \langle\vec\psi_{T}~|~ \vec\psi_{T} \rangle=  |\tilde t|^2  \neq \id.
\end{eqnarray}
Thus, the separate creation operators cannot be ``bosonic creation operators''. This is presented in the literature as a major problem that needs to be corrected. 
\nll
 In order to analyze this question we first make the link with the notation that is used conventionally in most of the literature, like e.g. in \cite{Mandel1995} p.511:
 \begin{eqnarray} 
 \hat B_{\vec \psi_{i}}  \rightarrow  \hat a_1,\qquad
  \hat B_{ \vec\psi_{R}}  \rightarrow  \hat a_2,\qquad
    \hat B_{ \vec\psi_{T} }  \rightarrow  \hat a_3.  
 \end{eqnarray}
\vskip 4ex
\begin{minipage}[t]{0.3\textwidth}
\begin{tikzpicture} [scale=3]
\draw [line width=7pt,gray] (-0.415,-0.4) --  (0.385,0.4);
\draw [->,green,dashed](0,-1) --  (0,0);
\draw [->,green,dashed](0,-1) --  (0,-0.5);\draw [->,green,dashed](0,-0.5) --  (0,0);
\draw [->,green,dashed](0,0) --  (0,1);
\draw [->,blue](-0.03,0.03) --  (-0.03,1);
\draw [->,blue](-1,0.03) --  (-0.5,0.03); \draw [->,blue](-0.5,0.03) --  (-0.03,0.03);
\draw [->,green,dashed](0,0) --  (1,0);
\draw [->,blue](-0.03,0.03) --  (1,0.03);
\draw [line width=3pt]  (0.2,1.1) -- (-0.2,1.1)   arc(180:0:0.195) ;
\draw [line width=3pt]  (1.1,0.2) -- (1.1,-0.2)   arc(-90:90:0.195) ;
\draw [domain=-0.8:-0.2,blue] plot (\x,{0.1+0.3*sin((\x+0.8)*pi/(-0.2+0.8) r)^4});
 \node at (-0.5,0.5) {$ {\color{blue} \vec\psi_{i} }$ };
\draw [domain=-0.8:-0.2,green,dashed] plot ({0.1+0.3*sin((\x+0.8)*pi/(-0.2+0.8) r)^4}, \x);
\node at (0.55,-0.5) {$ {\color{green} \vec\psi_{vac,i}  }$ };
\node at (0.1,1.0) {$y$};  \node at (1.0,0.1) {$x$};
\node at (0.35,1.2) {$D_Y$};  \node at (1.42,0.2) {$D_X$};
\node at (-0.8,1.2) {(a)};
\end{tikzpicture}
\end{minipage}
\hskip15ex
\begin{minipage}[t]{0.55\textwidth}
\begin{tikzpicture} [scale=3]
\draw [line width=7pt,gray] (-0.415,-0.4) --  (0.385,0.4);
\draw [->,green,dashed](0,-1) --  (0,0);
\draw [->,green,dashed](0,-1) --  (0,-0.5);\draw [->,green,dashed](0,-0.5) --  (0,0);
\draw [->,green,dashed](0,0) --  (0,1);
\draw [->,blue](-0.03,0.03) --  (-0.03,1);
\draw [->,blue](-0.8,0.03) --  (-0.5,0.03); \draw [->,blue](-0.5,0.03) --  (-0.03,0.03);
\draw [->,green,dashed](0,0) --  (1,0);
\draw [->,blue](-0.03,0.03) --  (1,0.03);
\draw [line width=3pt]  (0.2,1.1) -- (-0.2,1.1)   arc(180:0:0.195) ;
\draw [line width=3pt]  (1.1,0.2) -- (1.1,-0.2)   arc(-90:90:0.195) ;
\draw [domain=0.2:0.8,blue] plot (\x,{0.1+0.15*sin((\x-0.2)*pi/(0.8-0.2) r)^4});
 \node at (0.8,0.3) {$ {\color{blue} \vec\psi_{T} }$ };
\draw [domain=0.2:0.8,green,dashed] plot (\x,{-0.2+0.15*sin((\x-0.2)*pi/(0.8-0.2) r)^4});
 \node at (0.8,-0.3) {$ {\color{green} \vec\psi_{vac,R} }$ };
\draw [domain=0.2:0.8,green,dashed] plot ({0.1+0.15*sin((\x-0.2)*pi/(0.8-0.2) r)^4}, \x);
\node at (0.3,0.8) {$ {\color{green} \vec\psi_{vac,T} }$ };
\draw [domain=0.2:0.8,blue] plot ({-0.1-0.15*sin((\x-0.2)*pi/(0.8-0.2) r)^4}, \x);
\node at (-0.3,0.8) {$ {\color{blue} \vec\psi_{R} }$ };
\node at (0.1,1.0) {$y$};  \node at (1.0,0.1) {$x$};
\node at (0.35,1.2) {$D_Y$};  \node at (1.42,0.2) {$D_X$};
\node at (-0.7,1.2) {(b)}; 
\end{tikzpicture}
\end{minipage}
\vskip 1ex
\noindent
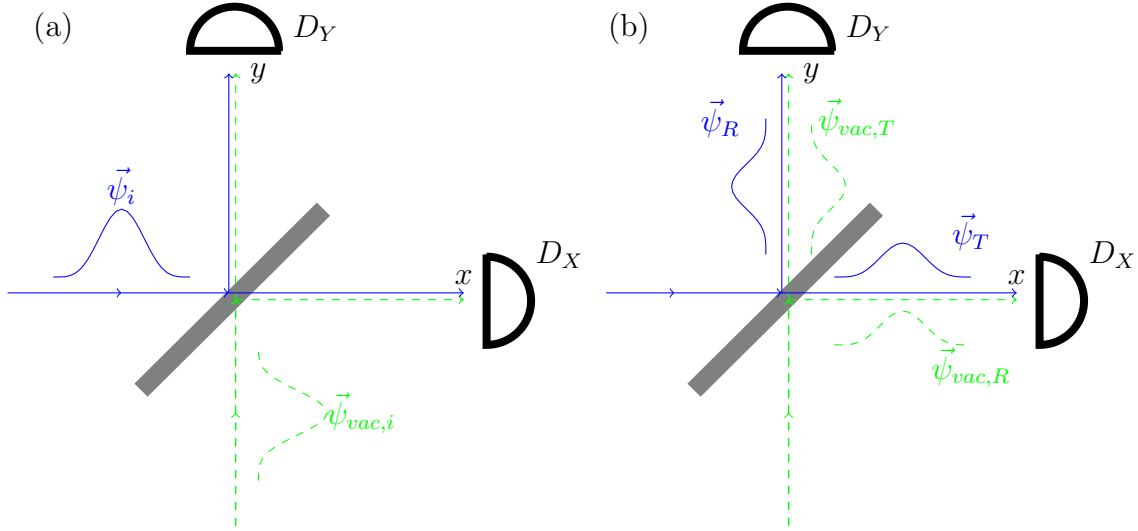
Figure 2. Schematic illustration of the description of beam splitters which uses a fourth port representing the ``incoming fluctuations of the vacuum''. (a) incoming pulses representing the initial state (in blue) and the vacuum fluctuations (dashed green curve) (b) reflection and transmission of  the initial pulse and the fluctuation pulse.
Fluctuation pulses are drawn in dashed green lines since they don't have any meaning neither in the classical nor in the quantum theory.
\vskip 4ex
The proposed correction is to add a fourth port represented by an operator $\hat a_0$ that should correspond to ``incoming fluctuations of the vacuum'', represented by green dashed lines in  Figure 2 (see e.g. \cite{Mandel1995} p.511 and p.640, \cite{Vogel2006} p.8, 229, \cite{Leonhardt2005} p.70, \cite{Leonhardt2010} p.96, 123, \cite{Bachor2019} p.102, \cite{Gerry2004} p.138, \cite{Grynberg2010} p.432,
\cite{Chiao2014} p.249 and footnote 1).
The addition of this port is supposed to solve the problem, since instead of 
 \begin{eqnarray} 
\hat a_2 &  =  & \tilde t~ \hat  a_1 \\
\hat a_3 & = &   \tilde r ~\hat  a_1, 
 \end{eqnarray}
its inclusion allows to construct modified operators
 \begin{eqnarray} 
\hat a'_2 &  :=  & \tilde t~ \hat  a_1 +  \tilde r~ \hat  a_0\\
\hat a'_3 & := &   \tilde r ~\hat  a_1 +  \tilde t~ \hat  a_0
 \end{eqnarray}
that satisfy the bosonic commutation relations
 \begin{eqnarray} 
\left[ \hat a'_2 ~,~ \hat a^{\prime\dag}_2   \right] &=& \id    \\
\left[ \hat a'_3 ~,~ \hat a^{\prime\dag}_3   \right] &=& \id
 \end{eqnarray}
 since $|\tilde r|^2+|\tilde t|^2 =1 $.
\nl

Although formally this seems to solve the ``problem'', yielding operators with apparently bosonic commutation relations, we have two criticisms of this construction: \\

(a) It is not a problem that the operators in \eqref{separateCR} do not satisfy the commutation relation for bosons. The photon is not an excitation of the individual transmitted pulse nor of the reflected one.
 The photon is an excitation of the global mode $ \vec\psi_{R}  +  \vec\psi_{T}$  and thus the bosonic nature of the photon is expressed by the global commutation relation \eqref{globalCR}.  So in fact the commutation relations  \eqref{separateCR} are correct, and they do not pose any problem, since creation operators in Fock space are also well-defined for unnormalized modes $\vec \psi$.
\\

(b) The formal expression $\hat  a^\dag_0$ that would have to correspond to $\hat B^\dag_{vac}$ does not have any well defined sense in the quantum field theory of light. There is no creation operator that could create ``fluctuations of the vacuum''.  In Figure 2, the pulses drawn in dashed lines, which are supposed to refer to vacuum fluctuations, don't have any meaning. Creation operators can only add photons carried by classical modes, to the vacuum or to other states.  
\\

Our main conclusion from the analysis presented above is that vacuum fluctuations don't play any role in the transmission of a one-photon pulse through a beam splitter.
\\
Our explanation for the discrepancy with much of the literature is that the notation $\hat a_1,\hat a_2, \hat a_3,\hat a_0$ is misleading and clearly insufficient for the description of the transmission of a one-photon pulse through a beam splitter. One difficulty is that the operator \eqref{separateB}, 
 $   \hat B^\dag_{ \vec\psi_{R}  +  \vec\psi_{T} }  $ has no natural representation in the conventional notation: it could maybe be noted  as $\hat a^\dag_{2+3}$, but it is not usually done. The notation also suggests that all operators that are denoted with $\hat a_j$ should satisfy the standard bosonic commutation relations, which is not well suited for the beam splitter, as shown by Eqs. \eqref{separateCR}.
 \nl
 Another shortcoming of this notation is that it does not include the propagation in time: When one writes $\hat a_3= \tilde t \hat  a_1 $,  the  operator $\hat  a_1$ is not the incoming one, but implicitly it must be interpreted as the time evolution of the incoming one (which is not the evolution in the Heisenberg representation, as we remarked in 
 \eqref{notHeisenberg} \eqref{Heisenberg}).
 \nl
  All this difficulties disappear if one uses the more precise notation $\hat B^\dag_{\vec \psi}$, which indicates on which classical mode $\vec \psi$ a photon is created. With this notation it is immediately clear that it does not make sense to define a creation operator like $\hat a^\dag_0$ that would create ``fluctuations of the vacuum'', since there is no $\vec \psi$ that can do that.
\\

The above remarks do not mean that there are no quantum fluctuations in the one-photon pulse transmission through a beam splitter. The detector readings have quantum fluctuations that can be traced back to the fact that the state of the optical quantum field is not an eigenstate of the observable measured by the detector. Our point is only that  the statement that ``vacuum fluctuations enter through the unused port'' is not well-defined within the theory and  that it is unnecessary, since, without any such addition, the  theory already gives a complete and consistent description of the process, including the fluctuations that will be manifest at the detectors. The fluctuations are a global feature of the quantum system. They are everywhere and delocalized. They do not enter through any particular port.
The photon states and in particular the vacuum are global entities, they are not localized in any specific port. If the electromagnetic field is in a one-photon state, one cannot claim that it is in the vacuum state anywhere, in particular not in an ``unused port''. The fluctuations that may be measured in a particular detector, will be the ones determined by the one-photon state, and not by some partially localized vacuum.
\\

As a final remark, in fact the model with four ports can describe a concrete physical situation, but only when an actual classical mode is injected into the fourth  port, as we will see in the next section on the Hong-Ou-Mandel effect (HOM) \cite{Hong1987} \cite{Bouchard2020}.

\section{The Hong-Ou-Mandel effect }

We consider two pulses 
 arriving on a 50-50\% beam splitter at $45^o$. 
$\vec\psi_{A,i}$ moves horizontally in the $x$  direction and $\vec\psi_{B,i}$ moves vertically in the $y$.
\vskip 4ex
\begin{minipage}[t]{0.3\textwidth}
\begin{tikzpicture} [scale=3]
\draw [line width=7pt,gray] (-0.415,-0.4) --  (0.385,0.4);
\draw [->,red](0,-1) --  (0,0);
\draw [->,red](0,-1) --  (0,-0.5);\draw [->,red](0,-0.5) --  (0,0);
\draw [->,red](0,0) --  (0,1);
\draw [->,blue](-0.03,0.03) --  (-0.03,1);
\draw [->,blue](-1,0.03) --  (-0.5,0.03); \draw [->,blue](-0.5,0.03) --  (-0.03,0.03);
\draw [->,red](0,0) --  (1,0);
\draw [->,blue](-0.03,0.03) --  (1,0.03);
\draw [line width=3pt]  (0.2,1.1) -- (-0.2,1.1)   arc(180:0:0.195) ;
\draw [line width=3pt]  (1.1,0.2) -- (1.1,-0.2)   arc(-90:90:0.195) ;
\draw [domain=-0.8:-0.2,blue] plot (\x,{0.1+0.3*sin((\x+0.8)*pi/(-0.2+0.8) r)^4});
 \node at (-0.5,0.5) {$ {\color{blue} \vec\psi_{A,i} }$ };
\draw [domain=-0.8:-0.2,red] plot ({0.1+0.3*sin((\x+0.8)*pi/(-0.2+0.8) r)^4}, \x);
\node at (0.55,-0.5) {$ {\color{red} \vec\psi_{B,i} }$ };
\node at (0.1,1.0) {$y$};  \node at (1.0,0.1) {$x$};
\node at (0.35,1.2) {$D_Y$};  \node at (1.42,0.2) {$D_X$};
\node at (0.35,1.2) {$D_Y$};  \node at (1.42,0.2) {$D_X$};
\node at (-0.8,1.2) {(a)}; 
\end{tikzpicture}
\end{minipage}
\hskip15ex
\begin{minipage}[t]{0.55\textwidth}
\begin{tikzpicture} [scale=3]
\draw [line width=7pt,gray] (-0.415,-0.4) --  (0.385,0.4);
\draw [->,red](0,-1) --  (0,0);
\draw [->,red](0,-1) --  (0,-0.5);\draw [->,red](0,-0.5) --  (0,0);
\draw [->,red](0,0) --  (0,1);
\draw [->,blue](-0.03,0.03) --  (-0.03,1);
\draw [->,blue](-0.8,0.03) --  (-0.5,0.03); \draw [->,blue](-0.5,0.03) --  (-0.03,0.03);
\draw [->,red](0,0) --  (1,0);
\draw [->,blue](-0.03,0.03) --  (1,0.03);
\draw [line width=3pt]  (0.2,1.1) -- (-0.2,1.1)   arc(180:0:0.195) ;
\draw [line width=3pt]  (1.1,0.2) -- (1.1,-0.2)   arc(-90:90:0.195) ;
\draw [domain=0.2:0.8,blue] plot (\x,{0.1+0.15*sin((\x-0.2)*pi/(0.8-0.2) r)^4});
 \node at (0.8,0.3) {$ {\color{blue} \vec\psi_{A,T} }$ };
\draw [domain=0.2:0.8,red] plot (\x,{-0.2+0.15*sin((\x-0.2)*pi/(0.8-0.2) r)^4});
 \node at (0.8,-0.3) {$ {\color{red} \vec\psi_{B,R} }$ };
\draw [domain=0.2:0.8,red] plot ({0.1+0.15*sin((\x-0.2)*pi/(0.8-0.2) r)^4}, \x);
\node at (0.3,0.8) {$ {\color{red} \vec\psi_{B,T} }$ };
\draw [domain=0.2:0.8,blue] plot ({-0.1-0.15*sin((\x-0.2)*pi/(0.8-0.2) r)^4}, \x);
\node at (-0.3,0.8) {$ {\color{blue} \vec\psi_{A,R} }$ };
\node at (0.1,1.0) {$y$};  \node at (1.0,0.1) {$x$};
\node at (0.35,1.2) {$D_Y$};  \node at (1.42,0.2) {$D_X$};
\node at (0.35,1.2) {$D_Y$};  \node at (1.42,0.2) {$D_X$};
\node at (-0.7,1.2) {(b)}; 
\end{tikzpicture}
\end{minipage}
\vskip 1ex
\noindent
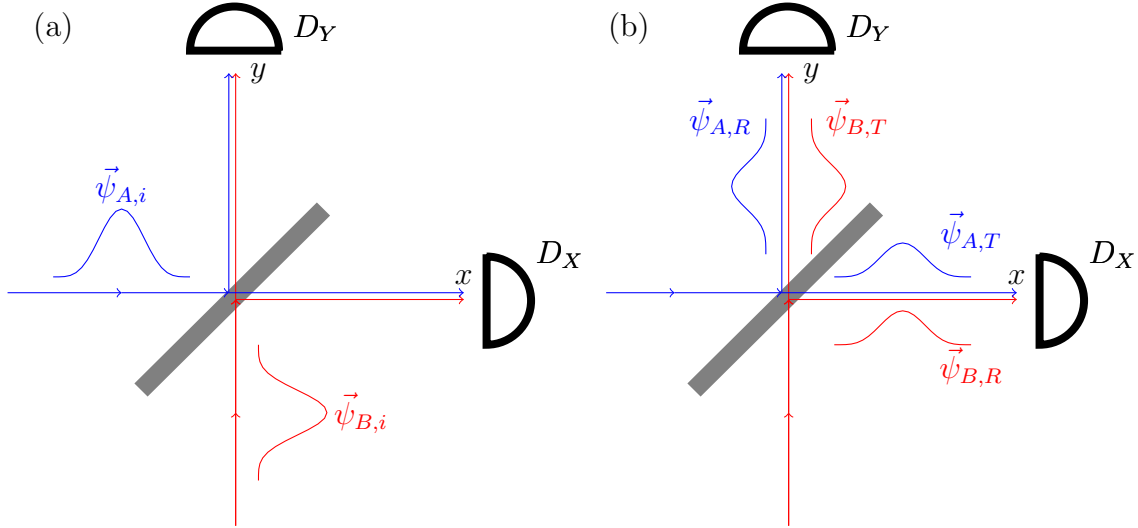
Figure 3. Schematic illustration of the Hong-Ou-Mandel effect. (a) Two disjoint identical pulses impinge on the beam splitter at the same time (b) each of them produces reflected and transmitted pulses. $D_X$ and $D_Y$ represent detectors, located at the same distance from the beam splitter.

\vskip 4ex
\subsection{Description of the initial classical modes   and their time evolution}
Each pulse at the initial time $t_i$ (before the pulses arrive on the beam splitter) is a classical mode. 
We will use the 
 {\it pulse shape function} $\vec S( r_1,r_2,r_3;k,t) $ defined in Eq. \eqref{pulse-shape}. For the pulse  envelope $\mathcal{E}(r_1)$ we have taken a $\sin^2$ with a finite support, so that there is no ambiguity  about when the process starts.
\nll
The initial modes $A$ and $B$ are chosen as
\begin{eqnarray}  \label{SA}
\vec\psi_{A,i}(\vec x) &=  &  \vec S(x,y,z;~ k_A, t=0) =:  \vec S_A(x,t=0)\\ \label{SB}
\vec\psi_{B,i}(\vec x) &= & \vec S(y,x,z;~ k_B, t=0)=:  \vec S_B(y,t=0),
\end{eqnarray}
one is oriented in the $x$ direction and the other one in the $y$ direction. 
 In \eqref{SA}\eqref{SB} we introduce an abridged notation $\vec S_A, \vec S_B$, indicating only the first spatial argument and the time argument.
\nll
After  crossing  the beam splitter,  the classical modes evolve into  a reflected pulse and a transmitted pulse:
\begin{eqnarray}
 \vec\psi_{A,i} &\rightarrow & \vec\psi_{A,R}  +  \vec\psi_{A,T}   \\
 \vec\psi_{B,i} &\rightarrow & \vec\psi_{B,R}  +  \vec\psi_{B,T}  
\end{eqnarray}
\begin{eqnarray}
\vec\psi_{A,R}(\vec x,t) & = & \tilde  r~  \vec S_A(y-ct,t) =: \tilde  r~   \vec Y_A(y) \\
 \vec\psi_{A,T}(\vec x,t) & = &  \tilde t~   \vec S_A(x-ct,t) =: \tilde t~ \vec X_A(x) \\
 \vec\psi_{B,R}(\vec x,t) & = & \tilde r~   \vec S_B(x-ct,t  =: \tilde  r~ \vec X_B(x)\\
 \vec\psi_{B,T} (\vec x,t) & = & \tilde t~   \vec S_B(y-ct,t) =: \tilde t~ \vec Y_B(y), 
\end{eqnarray}
where  $ \tilde r, \tilde t$ are the reflection and transmission coefficients, which in general satisfy  the relations  $ |\tilde r|^2+  |\tilde t|^2=1$ and 
 $ \tilde r^*  \tilde t+ \tilde r  \tilde t^*=0$.  For a  50-50\% beam splitter they satisfy  furthermore 
 \begin{eqnarray}
\tilde t=i  \tilde r, \qquad  {\rm i.e}\quad   \tilde r^2+  \tilde t^2=0.
\end{eqnarray}
We again introduce the abridged notations $ \vec Y_A (y), \vec X_A(x)$ $ \vec Y_B (y), \vec X_B(x)$ to improve the readability  of the construction below. ($\vec X$: propagation along the $x$-axis; $\vec Y$: propagation along the $y$-axis).

\subsection{Quantum dynamics of the two  photon state }

We assume that the system is prepared with one photon on each mode. The  two-photon state  in the Fock space is
\begin{eqnarray} \label{initstate} \nonumber
 | \Phi_i\rangle &=&\hat B^\dag_{\vec \psi_{A,i}}   \hat B^\dag_{\vec \psi_{B,i}}  |\varnothing\rangle \\
&=& \sqrt{2}|  \vec\psi_{A,i} \otimes_S \vec\psi_{B,i}  \rangle = | \vec\psi_{A,i} \otimes \vec\psi_{B,i} + \vec\psi_{B,i} \otimes \vec\psi_{A,i}   \rangle/\sqrt{2},
\end{eqnarray}
The mode functions must be normalized:
\begin{eqnarray} 
\int d^3 x   |\vec \psi_{A,i}(\vec x)|^2 =1, \qquad \int d^3 x   |\vec \psi_{B,i}(\vec x)|^2 =1,
\end{eqnarray}
which entails 
\begin{eqnarray} \nn
\left[\hat B_{\vec \psi_{A,i}}~,~ \hat B^\dag_{\vec \psi_{A,i}}  \right]&=&\id, \qquad\left[ \hat B_{\vec \psi_{B,i}}~,~ \hat B^\dag_{\vec \psi_{B,i}} \right]=\id, \\
\left[\hat B_{\vec \psi_{A,i}}~,~ \hat B^\dag_{\vec \psi_{B,i}} \right]&=&   \langle \vec \psi_{A,i}  |  \vec \psi_{B,i}\rangle =
\int d^3 x~  \vec \psi_{A,i}^* (\vec x)\cdot \vec \psi_{B,i}(\vec x)=0,
\end{eqnarray}
since the support on the two classical modes $ \vec \psi_{A,i}(\vec x)$ and $ \vec \psi_{B,i}(\vec x)$ is disjoint. This implies that the initial state \eqref{initstate} is normalized, $\langle  \Phi_i  | \Phi_i\rangle=1$.
\nll
This two-photon initial state evolves  to 
\begin{eqnarray} \nonumber
 && \vec\psi_{A,i} \otimes_S \vec\psi_{B,i}  \rightarrow 
 \left( \vec\psi_{A,R}  +  \vec\psi_{A,T} \right)  \otimes_S \left( \vec\psi_{B,R}  +  \vec\psi_{B,T}    \right)     \\
&= &  \vec\psi_{A,R}  \otimes_S  \vec\psi_{B,R}  +  \vec\psi_{A,R}  \otimes  \vec\psi_{B,T}  
 +\vec\psi_{A,T}   \otimes _S \vec\psi_{B,R}   + \vec\psi_{A,T} \otimes_S  \vec\psi_{B,T}    \\
 &= &  \tilde r   \vec Y_A \otimes_S  \tilde r \vec X_B +   \tilde r  \vec Y_A \otimes_S    \tilde t  \vec Y_B +
   \tilde t \vec X_A \otimes_S    \tilde r \vec X_B +   \tilde t \vec X_A \otimes_S    \tilde t  \vec Y_B \\    \nn
    &= &\frac{1}{2} \Bigl[\phantom{~} 
     \tilde r^2  \vec Y_A \otimes \vec X_B    +   \tilde t^2 \vec X_A \otimes      \vec Y_B
     +   \tilde r  \tilde t \left(  \vec Y_A \otimes     \vec Y_B + \vec X_A \otimes    \vec X_B  \right)  \\     \label{lastboson}
    & \phantom{=} & +~
          \tilde r^2 \vec X_B \otimes  \vec Y_A  +   \tilde t^2  \vec Y_B \otimes     \vec X_A
     +   \tilde r  \tilde t \left(  \vec Y_B \otimes     \vec Y_A + \vec X_B \otimes    \vec X_A  \right) \Bigr]. 
\end{eqnarray}
In the degenerate Hong-Ou-Mandel effect the shapes of the two incoming classical modes are the same and they arrive at the same time at the beam splitter and at the detectors,  which means that 
$  \vec Y_A=  \vec Y_B =: \vec Y$ and $ \vec X_A= \vec X_B =:\vec X$. Therefore the above expression can be rewritten as
\begin{eqnarray}  \label{HOMzero}
\Phi_f    &= & \frac{\tilde r^2+\tilde t^2}{\sqrt{2}} \left( \vec X \otimes  \vec Y + \vec Y \otimes \vec X  \right)
     +  \sqrt{2}~ \tilde r  \tilde t \left(  \vec Y \otimes     \vec Y +
\vec X \otimes    \vec X  \right). 
\end{eqnarray}
Since for a  50-50\% beam splitter $  \tilde r^2+  \tilde t^2=0$,  the final state for   the degenerate Hong-Ou-Mandel effect can be written as
\begin{eqnarray}  \label{HOM}
&& \Phi_i:= \sqrt{2}\left( \vec\psi_{A,i} \otimes_S \vec\psi_{B,i} \right) \rightarrow  
\Phi_f= \sqrt{2}~\tilde r  \tilde t \left(  \vec Y \otimes     \vec Y +  \vec X \otimes    \vec X  \right).
\end{eqnarray}

From the expression  \eqref{HOM} one can conclude that there will be no simultaneous detection of one photon in each detector, since the state does not contain terms of the form $\left( \vec X \otimes  \vec Y + \vec Y \otimes \vec X  \right)$. In order to make this statement more precise we have to construct a model for the detectors, i.e. we have to write the observables that correspond to single and double detections. A simple model for these observables can be constructed using the classical modes $\vec X$ and  $ \vec Y$: 
\nll
The observable corresponding to the detection of a single photon in the considered mode shape in the detector $D_X$ is 
\begin{eqnarray}
\hat{O}_{1X}  :=  | \vec X \rangle\langle \vec X | \otimes \id +  \id \otimes  | \vec X \rangle\langle \vec X |,
    \end{eqnarray}
    and correspondingly in the detector  $D_Y$
\begin{eqnarray}
 \hat{O}_{1Y}  :=   | \vec Y\rangle\langle \vec Y | \otimes \id +  \id \otimes  | \vec Y\rangle\langle \vec Y |.
\end{eqnarray}
The observable  corresponding to the detection of two photons in $D_X$ is
\begin{eqnarray}
\hat{O}_{2XX}  :=       | \vec X \rangle\langle \vec X | \otimes  | \vec X \rangle\langle \vec X |,
\end{eqnarray}
and correspondingly in the detector  $D_Y$
\begin{eqnarray}
 \hat{O}_{2\vec Y\vec Y}  :=       | \vec Y \rangle\langle \vec Y | \otimes  | \vec Y \rangle\langle \vec Y |.
\end{eqnarray}
The observable  corresponding to the detection of one photon in $D_X$ and simultaneously  one photon in $D_Y$ is
\begin{eqnarray}
\hat O_{2XY}  :=   | \vec Y \rangle\langle \vec Y | \otimes  | \vec X \rangle\langle \vec X | +     | \vec X \rangle\langle \vec X | \otimes  | \vec Y \rangle\langle \vec Y |.
\end{eqnarray}
Since $\hat O_{2XY}$ is a projector, the probability to observe simultaneously one photon in each detector in the final state $|\Phi_f\rangle$ is given by
\begin{eqnarray}
   \hbox{P}(D_X ~{\rm and~} D_Y) = \langle \Phi_f |  \hat O_{2XY}    |\Phi_f \rangle.
\end{eqnarray}
Inserting the expression we obtained for $|\Phi_f\rangle$ we get
\begin{eqnarray} 
\kern-50pt &&   \hbox{P}(D_X ~{\rm and~}  D_Y)   = \langle \Phi_f |  \hat O_{2XY}    |\Phi_f \rangle \\
\kern-50pt  &&=  2  | \tilde r|^2  |\tilde t |^2 \Big\langle \left(  \vec Y \otimes     \vec Y +  \vec X \otimes    \vec X  \right) \Big|~ \left[ | \vec Y \rangle\langle \vec Y | \otimes  | \vec X \rangle\langle \vec X |  
    \right]  ~ \Big|\left(  \vec Y \otimes     \vec Y +  \vec X \otimes    \vec X  \right)  \Big\rangle 
\end{eqnarray}
and since 
\begin{eqnarray}
 \left[ | \vec Y \rangle\langle \vec Y | \otimes  | \vec X \rangle\langle \vec X |  
    \right]  ~ | \vec Y \otimes     \vec Y  \rangle     &=& 0          \\  
     \left[ | \vec Y \rangle\langle \vec Y | \otimes  | \vec X \rangle\langle \vec X |  
    \right]  ~ | \vec X \otimes    \vec X  \rangle     &=& 0
\end{eqnarray}
we conclude that 
\begin{eqnarray}
 \hbox{P}(D_X ~{\rm and~}  D_Y) = 0,
\end{eqnarray}
i.e. the probability for simultaneous detection of one photon in each detector is zero, which is the main characteristic of the Hong-Ou-Mandel effect.
\nl
\subsection{Photons are not like classical waves nor like classical particles}

We emphasize that the last equality of \eqref{lastboson}, and thus \eqref{HOMzero} and \eqref{HOM}, are only true because the photons are bosons (i.e. indistinguishable quanta or ``particles''). The effect of the bosonic symmetrization is essential for the HOM effect.
\nl

In particular, as it is well known, that the HOM effect is an exclusively quantum effect, that does not happen with classical waves. Indeed the classical waves arrive at the two detectors simultaneously, and thus it is excluded that only one of the detectors is activated. The events that  for photons have probability zero, for classical waves would have probability one. 
 \nl
 
The behavior of photons in the HOM effect is also completely different from that of classical particles. Indeed, for classical particles having $1/2$ probability of transmission and $1/2$ of reflection, the probability of detecting two particles in the detector $ D_X$ would be $1/4$,  and the same $1/4$  for detector $ D_Y$. The probability to detect one particle in each detector would be $1/2$. Indeed, denoting by  $ D_X, D_Y$
the horizontal and the vertical detectors, and labeling the particles moving initially horizontally and vertically by $X_i$ and $Y_i$ respectively, 
the classical probabilities are
\begin{eqnarray} 
\hbox{P}({\rm 2~particles~ in ~}  D_X)& = & \hbox{P}({\rm 1~ particle ~in ~}  D_X)\times \hbox{P}({\rm 1~ particle~ in ~}  D_X)=\frac{1}{2}\frac{1}{2}=\frac{1}{4}
 \\ 
\hbox{P}({\rm 2 ~particles~ in~ } D_Y)& = & \hbox{P}({\rm 1~ particle~ in~ } D_Y)\times \hbox{P}({\rm 1~ particle~ in ~} D_Y)=\frac{1}{2}\frac{1}{2}=\frac{1}{4} ~~~~~~~~~
\end{eqnarray}
and \begin{eqnarray}
\nn
\hbox{P}({\rm 1~ particle~ in ~}  D_X~&&  \kern-15pt{\rm   and~1~particle~in~ } D_Y)       \\ \nn
 & = &  \hbox{P}( {\rm particle~Y}_i {\rm ~in ~}  D_Y {\rm ~and~ particle~ X}_i {\rm ~in~ } D_X)   \\ 
&  & +\hbox{P}( {\rm particle~ Y}_i {\rm ~ in~}  D_X {\rm ~and~ particle~X}_i {\rm~ in~}  D_Y ) \phantom{xxxxx}\\ 
 & = & \frac{1}{2}\frac{1}{2}+\frac{1}{2}\frac{1}{2}=\frac{1}{2},
\end{eqnarray}
i.e. the event that has zero probability for photons has probability $1/2$ for classical particles.
\nl

 This example gives an illustration for the question on whether photons are particles or waves, or both. The answer is that they are neither particles nor waves in the classical sense. It is not that photons ``behave sometimes like particles and sometimes like waves'', as it was often stated in the early stages of the development of quantum mechanics. They are purely quantum entities, that have properties that do not exist in classical objects.
\\  

\section{Remarks on the concept of photons}

In his recent book \cite{Shore2020}, Bruce W. Shore gave a detailed account of the historical evolution of the concept of photons, as well as the different interpretations that have been formulated in different contexts.
In the present article we apply tools of quantum field theory to some questions of quantum optics involving passive dielectrics,
 in order to highlight some aspects of the concept of photons.
From the theoretical point of view, in the framework of quantum field theory, photons are the quanta of the electromagnetic field, that are created by the bosonic  operators  $\hat B^\dag_{\vec \psi}$ carried by a classical mode $\vec \psi$  (or electromagnetic configuration), applied on the vacuum state in Fock space. They are bosons, which means that they are intrinsically indistinguishable and their states are symmetric with respect to permutations.
In the context of the Hong-Ou-Mandel effect one often uses the expression ``two indistinguishable photons'' with a slightly different meaning, which is to be interpreted as ``a two-photon state, carried by two classical modes (or electromagnetic configurations) that are disjoint  and have the same shape (carrier frequency, pulse shape and polarization) and arrive simultaneously on the beam splitter''.
\nl

We can further remark that ``two photons''  does not mean ``one photon and another  photon''. Two photons means a two-photon state in the bosonic Fock space. Because of the bosonic nature, expressed by the symmetrization of the states, two photons are not the juxtaposition of one photon and another photon. A two-photon state  is a global entity that has properties that don't follow from the properties of single-photon states.
The HOM effect is an illustrative example of this fact. If two photons were the juxtaposition of one photon and another photon the HOM effect would not exist.
\nl

We remark furthermore that in the case of a single-photon pulse going through a beam splitter, discussed in Section \ref{Section-one-photon-beam splitter}, once the pulse has gone through the beam splitter, the single-photon state is both in the transmitted and in the reflected pulses. Neither of them can be called individually a photon.

\begin{appendices}
\section{Relation between the position Fock space   and the   reciprocal Fock space 
\label{Appendix-reciprocal}}

The description of photon states that we have used is defined on a {\it position space} representation of the Hilbert space $\mathcal{H}_{cl}$. In the literature, in particular for the quantized free fields, a {\it reciprocal space} representation is often used \cite{Cohen-Tannoudji1989}\cite{Spohn2004}. In this Appendix we describe the relation between the two representations. We first remark that in quantum mechanics one has to construct together the 
Hilbert space of states a representation and the operators describing the observables in that representation. The mathematical form of the state space alone is not enough to provide the physical interpretation of the theory. As formulated by Dirac in the theory of transformations \cite{Dirac1981},
if two Hilbert spaces are related by a unitary transformation and the observables are transformed accordingly, then all the physical predictions are identical. The physically observable quantities appear through expectation values, involving both the states and the observables.
\nll
The reciprocal space of the Hilbert and Fock spaces can be defined as follows. We start with the position space representation $\mathcal{H}_{cl}$, and an orthonormal basis $\{ \vec \varphi_\kappa \}$ of eigenvectors of  the frequency operator $\Omega$. Then any $\vec\psi(\vec x)\in \mathcal{H}_{cl}$ can be developed in this basis as
\begin{eqnarray}
\vec\psi(\vec x) &= \int d\kappa~ z(\kappa)~ \vec \varphi_\kappa(\vec x), 
\qquad {\rm with~} z(\kappa)  = \int_{\RR^3}d^3x~ \vec \varphi^{\kern1pt*}_\kappa(\vec x) \cdot \vec\psi(\vec x) \in \CC.
 \end{eqnarray} 
  The coefficients $z(\kappa)$  are in the Hilbert space
 \begin{eqnarray}
\tilde{\mathcal{H}}_{cl} &:= \{ z(\kappa) ~|~ \langle z |  z \rangle_{\tilde{\mathcal{H}}_{cl}}  <\infty   \},
\end{eqnarray}
with the scalar product
\begin{eqnarray} \label{def-scalar-product-z}
 \langle z_A |  z_B \rangle_{ \tilde{\mathcal{H}}_{cl} } &:= \int d\kappa~ z_A^*(\kappa)~  z_B (\kappa).
\end{eqnarray}
The map 
\begin{eqnarray} 
\mathcal{I}:  \mathcal{H}_{cl}  & \to & \tilde{\mathcal{H}}_{cl} \\    \label{def-I}
\vec\psi(\vec x)  & \mapsto & \left[\mathcal{I} \vec\psi \right] (\kappa)= z(\kappa)  = \int_{\RR^3}d^3x~ \vec \varphi^{\kern1pt*}_\kappa(\vec x)\cdot \vec\psi(\vec x)
\end{eqnarray}
is a unitary isomorphism, since it is bijective and it preserves the scalar products,
\begin{eqnarray} 
\langle\vec \psi_A |  \vec \psi_B  \rangle = \langle \mathcal{I} \vec\psi_A |  \mathcal{I} \vec \psi_B  \rangle_{ \tilde{\mathcal{H}}_{cl}}.
\end{eqnarray}
This is a direct consequence of the fact that the basis  $\{ \vec \varphi_\kappa \}$ is orthonormal.
\nll
The isomorphism of the Hilbert spaces $\mathcal{H}_{cl}$ and  $\tilde{\mathcal{H}}_{cl} $ extends to the corresponding Fock spaces
${\F}^{\mathfrak{B}}(\mathcal{H}_{cl}) $ and $\tilde{{\F}}^{\mathfrak{B}}:={\F}^{\mathfrak{B}}(\tilde{\mathcal{H}}_{cl}) $, where
\begin{eqnarray}
\tilde{{\F}}^{\mathfrak{B}}&:=&
\oplus_{m=0}^{\infty}{\tilde\F}^{\mathfrak{B}}_m, \\
\tilde{\F}^{\mathfrak{B}}_0& := & \CC \\
\tilde{\F}^{\mathfrak{B}}_1 & := &  \tilde{\mathcal{H}}_{cl} \\
\tilde{\F}^{\mathfrak{B}}_2 & := &    \tilde{\mathcal{H}}_{cl} \otimes_S \tilde{\mathcal{H}}_{cl} \\
&\vdots & \nonumber
\end{eqnarray}
The operators acting on these two Hilbert spaces are related by
\begin{eqnarray} 
\hat {\tilde{O}} &=   \mathcal{I} \hat{O} \mathcal{I}^{-1}.
\end{eqnarray}
In particular, the creation-annihilation operators are related by 
\begin{eqnarray} 
\hat {\tilde{B}}^\dag_{z} = \hat {\tilde{B}}^\dag_{\mathcal{I}\vec \psi} &=   \mathcal{I} \hat{B}^\dag_{\vec\psi} \mathcal{I}^{-1}.
\end{eqnarray}
For instance,
\begin{eqnarray} 
\hat {\tilde{B}}^\dag_{z} |\tilde{\varnothing} \rangle &=   \mathcal{I} \hat{B}^\dag_{\vec\psi} \mathcal{I}^{-1} | \tilde{\varnothing} \rangle 
=   \mathcal{I} \hat{B}^\dag_{\vec\psi}  |\varnothing \rangle
=  \mathcal{I}   |\vec\psi\rangle = |  z \rangle =  |  \mathcal{I} \vec \psi\rangle
=\hat {\tilde{B}}^\dag_{\mathcal{I}\vec \psi} | \tilde{\varnothing} \rangle.
\end{eqnarray}
With this construction we can make the link between the position  Fock space representation of the electromagnetic observables and their reciprocal Fock space representation. For instance for the vector potential we have
in the position space representation
 \begin{eqnarray}
\sqrt{\epsilon} \vec {\hat A}=\vec {\hat A'}(\vec x)  & =  &\sqrt{\frac{\hbar}{2\epsilon_0}}  \int d\kappa~  
  \omega_\kappa^{-1/2} \left(  \vec\varphi_\kappa(\vec x)~\hat{B}_{\vec\varphi_\kappa}+ \vec\varphi^{\kern0.3ex *}_\kappa(\vec x)~\hat{B}_{\vec\varphi_\kappa}^\dag \right) 
\end{eqnarray}
and in the reciprocal space representation
 \begin{eqnarray}
\sqrt{\epsilon} \vec {\hat {\tilde A}}=\vec {\hat {\tilde A'}}(\vec x)  & = & \sqrt{\frac{\hbar}{2\epsilon_0}}  \int d\kappa~  
  \omega_\kappa^{-1/2} \left(  \vec\varphi_\kappa(\vec x)~\hat{{\tilde B}}_{\mathcal{I}\vec\varphi_\kappa}
  + \vec\varphi^{\kern0.3ex *}_\kappa(\vec x)~\hat{\tilde B}_{\mathcal{I}\vec\varphi_\kappa}^\dag \right). 
\end{eqnarray}
We remark that  $\left( \mathcal{I}\vec\varphi_\kappa\right) (\kappa') = \delta(\kappa'-\kappa)$,
where $\delta(\kappa'-\kappa)$ is a notation for combination of Dirac and Kronecker delta functions. This follows from the definition \eqref{def-I},
which applied to $\vec\psi = \vec\varphi_{\kappa'} $ yields
\begin{eqnarray}  
\left( \mathcal{I}\vec\varphi_{\kappa'}\right) (\kappa)  =  \int_{\RR^3}d^3x~ \vec \varphi^{\kern1pt*}_\kappa(\vec x)\cdot  \vec \varphi_{\kappa'}(\vec x) =\delta(\kappa-\kappa').
 \end{eqnarray} 
\nl

In order to make the link with the formulas used most often in the literature, we consider the free field case, i.e. $\epsilon(\vec x)=1$.
In this case
 \begin{eqnarray}
\vec\varphi_\kappa(\vec x) = \frac{1}{(2\pi)^{3/2}} \vec \varepsilon_s(\vec k) e^{i\vec k\cdot \vec x},
\end{eqnarray}
where $\vec k\in\RR^3$ and $s=\pm$ is an index for the two polarizations, i.e. $\kappa=(\vec k, s)$.
In the reciprocal space
 \begin{eqnarray}
\left( \mathcal{I}\vec\varphi_{\kappa'} \right) (\kappa) = \delta(\vec k-\vec k') \delta_{s,s'}
\end{eqnarray}
and the corresponding creation-annihilation operators are denoted
 \begin{eqnarray}
\hat a^\dag_{\vec k\kern1pt',s'} := \hat {\tilde{B}}^\dag_{\delta(\vec k-\vec k') \delta_{s,s'}}.
\end{eqnarray}
With this notation we can write
 \begin{equation} \label{eqAa}
 \phantom{xxxx}
 \vec {\hat {\tilde A}}=   \frac{1}{(2\pi)^{3/2}} \sqrt{\frac{\hbar}{2\epsilon_0}} \int d^3k \sum_{s=\pm}   
  \omega_{\vec k}^{-1/2} \left(  \vec \varepsilon_s(\vec k) e^{i\vec k\cdot \vec x}~\hat a_{\vec k,s} 
  +\vec \varepsilon_s^{\kern2pt*}(\vec k) e^{-i\vec k\cdot \vec x}~\hat a^\dag_{\vec k,s} \right), 
\end{equation}
with $\omega_{\vec k} =c |\vec k|$. This expression coincides with the usual formulas of the literature, like e.g. in \cite[p.483, (10.4-38)]{Mandel1995}\cite{Cohen-Tannoudji1989}, in the infinite volume limit.
\nll
We remark in conclusion that the Fock space that is mostly used in the free field case is the reciprocal one. For the space-time representation of the photons, in particular in the presence of a medium, we used the position space representation of the Fock space, which allows a spatial representation of the photons states, as in Eq. \eqref{one-photon-psi-x}. 
\section{Proof of the time evolution formulas  of N-photon   states 
\label{Appendix-proof-evolution-photon-states}}

We are going to prove the formulas \eqref{onepev},\eqref{twopev},\eqref{Npev} of Section  \ref{Section-evolution-photon-states}, i.e. 
that for an  N-photon initial condition
 $| \Phi(t=0)\rangle= \prod_{j=1}^N \hat B_{\vec\psi_j(t=0)}^\dag |\varnothing\rangle $ the time evolution is 
\begin{eqnarray} \label{Npev-11}
 | \Phi(t)\rangle= \prod_{j=1}^N \hat B_{\vec\psi_j(t)}^\dag |\varnothing\rangle, 
\end{eqnarray} 
where $\vec\psi_j(t)$  are the solutions of the classical equation \eqref{ClassicalWaveEq} for the corresponding initial conditions.
\nll 
{\bf Proof:}  Although  the proof  of the general N-photon case given  in  {\bf (c)} below implies  of course the results for the one- and two-photon cases, since it is notationally harder to read  we will first give the proofs for the simplest particular cases.
We will use the relation
\begin{eqnarray}  \label{B-linearity}
\hat B^\dag_{\alpha \vec\psi_A+\beta  \vec\psi_B} = \alpha \hat B^\dag_{ \vec\psi_A}  +\beta  \hat B^\dag_{ \vec\psi_B} \quad {\rm for ~} \alpha,\beta\in \CC.
\end{eqnarray}
We will write $\sum_\kappa$ instead of $\int d\kappa$ to simplify the notation.

{\bf  (a)} We first prove the statement for one-photon states, i.e. that the state defined by \eqref{onepev} satisfies 
\eqref{SchrodingerFockEq}. We start by expressing the time evolution of the classical modes in terms of the  eigenfunctions $\vec\varphi_\kappa$  of $\Omega$:
\begin{eqnarray} 
\vec \psi(t)& = &  \sum_\kappa  e^{-i\omega_\kappa t} \vec\varphi_\kappa \alpha_\kappa, \qquad \alpha_\kappa=\langle  \vec\varphi_\kappa |  \vec\psi(t=0)\rangle,
\end{eqnarray}
and
\begin{eqnarray}  
| \Phi(t) \rangle=\hat B_{\vec\psi(t)}^\dag|\varnothing\rangle=   \hat B^\dag_{ \sum_\kappa  e^{-i\omega_\kappa t} \vec\varphi_\kappa \alpha_\kappa} |\varnothing\rangle 
= \sum_\kappa  e^{-i\omega_\kappa t}  \alpha_\kappa \hat B^\dag_{\vec\varphi_\kappa}  |\varnothing\rangle. 
\end{eqnarray}
 With this representation we can write
\begin{eqnarray}  \label{dtPhi}
i\hbar\frac{\partial}{\partial t}  | \Phi(t)\rangle& = &    \sum_\kappa   \hbar \omega_\kappa ~e^{-i\omega_\kappa t}  \alpha_\kappa \hat B^\dag_{\vec\varphi_\kappa}  |\varnothing\rangle.
\end{eqnarray}
 Using the representation of the Hamiltonian as $\hat H=\sum_{\kappa'} \hbar\omega_{\kappa'} \hat B^\dag_{\vec\varphi_{\kappa'}}\hat B_{\vec\varphi_{\kappa'}}$ we can write
\begin{eqnarray} 
\hat H| \Phi(t)\rangle& = &  
\sum_{\kappa',\kappa} \hbar\omega_{\kappa'} e^{-i\omega_\kappa t}  \alpha_\kappa \hat B^\dag_{\vec\varphi_{\kappa'}}   \hat B_{\vec\varphi_{\kappa'}}
 \hat B^\dag_{\vec\varphi_\kappa}  |\varnothing\rangle \\
 & = &  
\sum_{\kappa',\kappa} \hbar\omega_{\kappa'} e^{-i\omega_\kappa t}  \alpha_\kappa \hat B^\dag_{\vec\varphi_{\kappa'}}
\left( \hat B^\dag_{\vec\varphi_{\kappa}}\hat B_{\vec\varphi_{\kappa'}} +\delta_{\kappa,\kappa'} \right)  |\varnothing\rangle \\
 & = &  \sum_{\kappa} \hbar\omega_{\kappa} e^{-i\omega_\kappa t}  \alpha_\kappa  \hat B^\dag_{\vec\varphi_\kappa}  |\varnothing\rangle 
=  i\hbar\frac{\partial}{\partial t}  | \Phi(t)\rangle,
\end{eqnarray}
where we have used the commutation relations
$\hat B_{\vec\varphi_{\kappa'}}\hat B^\dag_{\vec\varphi_{\kappa}} = \hat B^\dag_{\vec\varphi_{\kappa}}\hat B_{\vec\varphi_{\kappa'}} +\delta_{\kappa,\kappa'}$, the fact that  $\hat B_{\vec\varphi_{\kappa'}}|\varnothing\rangle =0$ and \eqref{dtPhi}, which completes the proof.
\nl

{\bf  (b)} Next we prove the relation for two-photon states, i.e. we prove  that the state defined by \eqref{twopev} satisfies \eqref{SchrodingerFockEq}.
First we expand in terms of the eigenfunction of $\Omega$:
\begin{eqnarray} 
\vec \psi_j(t)& = &  \sum_{\kappa_j}  e^{-i\omega_{\kappa_j} t} \vec\varphi_{\kappa_{j}} \alpha_{\kappa_{j}}, \qquad \alpha_{\kappa_j}=\langle  \vec\varphi_{\kappa_j} | \vec\psi(t=0)\rangle, \qquad j=A,B,
\end{eqnarray}
which allows us to write

\begin{eqnarray} 
 | \Phi(t)\rangle &=& \hat B_{\vec\psi_B(t)}^\dag \hat B_{\vec\psi_A(t)}^\dag|\varnothing\rangle 
=  \prod_{j=A,B}    \left( \sum_{\kappa_j} 
  e^{-i\omega_{\kappa_j}  t}  \alpha_{\kappa_j}  \hat B^\dag_{\vec\varphi_{\kappa_j} } \right) |\varnothing\rangle \\
 &=&   
   \left( \sum_{\kappa_A} e^{-i\omega_{\kappa_A}  t}  \alpha_{\kappa_A}  \hat B^\dag_{\vec\varphi_{\kappa_A} } \right)  \left( \sum_{\kappa_B} e^{-i\omega_{\kappa_B}  t}  \alpha_{\kappa_B}  \hat B^\dag_{\vec\varphi_{\kappa_B} } \right) |\varnothing\rangle \\  \label{phiBB}
    &=&   
 \sum_{\kappa_A,\kappa_B} e^{-i(\omega_{\kappa_A} + \omega_{\kappa_B} ) t} ~ \alpha_{\kappa_A}  \alpha_{\kappa_B}  \hat B^\dag_{\vec\varphi_{\kappa_A} }  \hat B^\dag_{\vec\varphi_{\kappa_B} }  |\varnothing\rangle  
\end{eqnarray} 
and 
\begin{eqnarray}\label{dttwo}
i\hbar\frac{\partial}{\partial t}  | \Phi(t)\rangle & = &
 \sum_{\kappa_A,\kappa_B} \hbar(\omega_{\kappa_A} + \omega_{\kappa_B} )~e^{-i(\omega_{\kappa_A} + \omega_{\kappa_B} ) t} ~ \alpha_{\kappa_A}  \alpha_{\kappa_B}  \hat B^\dag_{\vec\varphi_{\kappa_A} }  \hat B^\dag_{\vec\varphi_{\kappa_B} }  |\varnothing\rangle. 
\end{eqnarray} 
Applying the Hamiltonian to \eqref{phiBB} yields
\begin{eqnarray}  
\hat H   | \Phi(t)\rangle &=& 
\sum_{\kappa'} \hbar \omega_{\kappa'}  \hat B_{\vec\varphi_{\kappa'} }^\dag  \hat B_{\vec\varphi_{\kappa'} } ~
\sum_{\kappa_A,\kappa_B} e^{-i(\omega_{\kappa_A} + \omega_{\kappa_B} ) t} ~ \alpha_{\kappa_A}  \alpha_{\kappa_B}  \hat B^\dag_{\vec\varphi_{\kappa_A} }  \hat B^\dag_{\vec\varphi_{\kappa_B} }  |\varnothing\rangle  \\    \label{HBB}
 &=& 
 \sum_{\kappa_A,\kappa_B}  
 ~e^{-i(\omega_{\kappa_A} + \omega_{\kappa_B} ) t} ~ \alpha_{\kappa_A}  \alpha_{\kappa_B} ~
\sum_{\kappa'} \hbar\omega_{\kappa'}  \hat B^\dag_{\vec\varphi_{\kappa'}}    \hat B_{\vec\varphi_{\kappa'}}
 \hat B^\dag_{\vec\varphi_{\kappa_A} }  \hat B^\dag_{\vec\varphi_{\kappa_B} }  |\varnothing\rangle.
 \end{eqnarray} 
Using the commutation relations $\hat B_{\vec\varphi_{\kappa'}}\hat B^\dag_{\vec\varphi_{\kappa_j}} = \hat B^\dag_{\vec\varphi_{\kappa_j}}\hat B_{\vec\varphi_{\kappa'}} +\delta_{\kappa_j,\kappa'}$ and the fact that  $\hat B_{\vec\varphi_{\kappa'}}|\varnothing\rangle =0$ we can write
\begin{eqnarray} \label{recursion2}
 \hat B_{\vec\varphi_{\kappa'}} \hat B^\dag_{\vec\varphi_{\kappa_A} }  \hat B^\dag_{\vec\varphi_{\kappa_B} }  |\varnothing\rangle 
 &=& \delta_{\kappa',\kappa_A} \hat B^\dag_{\vec\varphi_{\kappa_B} }  |\varnothing\rangle 
 +   \hat B^\dag_{\vec\varphi_{\kappa_A} }  \hat B_{\vec\varphi_{\kappa'}} \hat B^\dag_{\vec\varphi_{\kappa_B} }  |\varnothing\rangle  \\ 
  &=&  \delta_{\kappa',\kappa_A} \hat B^\dag_{\vec\varphi_{\kappa_B} }  |\varnothing\rangle 
   +  \delta_{\kappa',\kappa_B}   \hat B^\dag_{\vec\varphi_{\kappa_A} }   |\varnothing\rangle  
    +   \hat B^\dag_{\vec\varphi_{\kappa_A} }  \hat B^\dag_{\vec\varphi_{\kappa_B} }  \hat B_{\vec\varphi_{\kappa'}}  |\varnothing\rangle  \\
      &=&  \delta_{\kappa',\kappa_A} \hat B^\dag_{\vec\varphi_{\kappa_B} }  |\varnothing\rangle 
   +  \delta_{\kappa',\kappa_B}   \hat B^\dag_{\vec\varphi_{\kappa_A} }   |\varnothing\rangle,  
  \end{eqnarray} 
and thus 
\begin{eqnarray}  
\sum_{\kappa'} \hbar\omega_{\kappa'}  \hat B^\dag_{\vec\varphi_{\kappa'}}    \hat B_{\vec\varphi_{\kappa'}}
 \hat B^\dag_{\vec\varphi_{\kappa_A} }  \hat B^\dag_{\vec\varphi_{\kappa_B} }  |\varnothing\rangle
                &=&  \sum_{\kappa'} \hbar\omega_{\kappa'}  \hat B^\dag_{\vec\varphi_{\kappa'}}  
   \left(     \delta_{\kappa',\kappa_A} \hat B^\dag_{\vec\varphi_{\kappa_B} }  |\varnothing\rangle 
   +  \delta_{\kappa',\kappa_B}   \hat B^\dag_{\vec\varphi_{\kappa_A} }   |\varnothing\rangle \right) ~~~~~~~~~~\\
       &=&   \hbar\omega_{\kappa_A}  \hat B^\dag_{\vec\varphi_{\kappa_A}}  
  \hat B^\dag_{\vec\varphi_{\kappa_B} }  |\varnothing\rangle 
   +  \hbar\omega_{\kappa_B}   \hat B^\dag_{\vec\varphi_{\kappa_B} }   \hat B^\dag_{\vec\varphi_{\kappa_A} }   |\varnothing\rangle  \\
     &=&   \left(\hbar\omega_{\kappa_A}   +  \hbar\omega_{\kappa_B}\right)  \hat B^\dag_{\vec\varphi_{\kappa_A}}  
  \hat B^\dag_{\vec\varphi_{\kappa_B} }  |\varnothing\rangle, 
\end{eqnarray} 
which inserted into \eqref{HBB} yields
\begin{eqnarray}  
 \nn
\hat H  | \Phi(t)\rangle &=& 
      \sum_{\kappa_A,\kappa_B} 
 ~e^{-i(\omega_{\kappa_A} + \omega_{\kappa_B} ) t} ~ \alpha_{\kappa_A}  \alpha_{\kappa_B} ~
\left(  \hbar\omega_{\kappa_A} 
   +  \hbar\omega_{\kappa_B} \right) \hat B^\dag_{\vec\varphi_{\kappa_A} }   \hat B^\dag_{\vec\varphi_{\kappa_B} }    |\varnothing\rangle  \\
    &=&       i\hbar\frac{\partial}{\partial t}  | \Phi(t)\rangle, 
 \end{eqnarray} 
where in the last equality we have used the relation \eqref{dttwo}, which completes the proof.
\nl

{\bf  (c)} We can generalize the preceding proof for  $N$-photon states as follows.
First we expand in terms of the eigenfunction of $\Omega$:
\begin{eqnarray}  
\vec \psi_j(t)& = &  \sum_{\kappa_j}  e^{-i\omega_{\kappa_j} t} \vec\varphi_{\kappa_{j}} \alpha_{\kappa_{j}}, \qquad \alpha_{\kappa_j}=\langle  \vec\varphi_{\kappa_j} | \vec\psi(t=0)\rangle, \qquad j=1,\ldots N,
\end{eqnarray}
which allows us to write
\begin{eqnarray} 
 | \Phi(t)\rangle &=& \prod_{j=1}^N \hat B_{\vec\psi_j(t)}^\dag|\varnothing\rangle 
=  \prod_{j=1}^N    \left( \sum_{\kappa_j} 
  e^{-i\omega_{\kappa_j}  t} ~ \alpha_{\kappa_j}  \hat B^\dag_{\vec\varphi_{\kappa_j} } \right) |\varnothing\rangle \\
 &=&   
   \left( \sum_{\kappa_1} e^{-i\omega_{\kappa_1}  t} ~ \alpha_{\kappa_1}  \hat B^\dag_{\vec\varphi_{\kappa_1} } \right) \ldots \left( \sum_{\kappa_N} e^{-i\omega_{\kappa_N}  t} ~ \alpha_{\kappa_N}  \hat B^\dag_{\vec\varphi_{\kappa_N} } \right) |\varnothing\rangle \\   \label{vvvv}
    &=&   
 \sum_{\kappa_1,\ldots,\kappa_N} e^{-i\sum_{j=1}^N\omega_{\kappa_j}  t} ~ \left(\prod_{j''}\alpha_{\kappa_{j''}}\right)  ~ \left(\prod_{j'}\hat B^\dag_{\vec\varphi_{\kappa_{j'}} } \right)   |\varnothing\rangle  
\end{eqnarray} 
and 
\begin{eqnarray}\label{dttwo11}
i\hbar\frac{\partial}{\partial t}  | \Phi(t)\rangle & = & 
 \sum_{\kappa_1,\ldots,\kappa_N} \left( \sum_{j}\hbar\omega_{\kappa_j} \right)e^{-i\sum_{j}\omega_{\kappa_j}  t} ~ \prod_{j''}\alpha_{\kappa_{j''}}  ~ \prod_{j'}\hat B^\dag_{\vec\varphi_{\kappa_{j'}} }   |\varnothing\rangle.  
\end{eqnarray} 
Applying the Hamiltonian to \eqref{vvvv} yields
\begin{align}  
\hat H  | \Phi(t)\rangle  &= 
\sum_{\kappa'} \hbar \omega_{\kappa'}  \hat B_{\vec\varphi_{\kappa'} }^\dag  \hat B_{\vec\varphi_{\kappa'} } ~
 \sum_{\kappa_1,\ldots,\kappa_N} e^{-i\sum_{j=1}^N\omega_{\kappa_j}  t} ~ \left(\prod_{j''}\alpha_{\kappa_{j''}}\right)  ~ \left(\prod_{j'}\hat B^\dag_{\vec\varphi_{\kappa_{j'}} } \right)   |\varnothing\rangle \\
  \label{vvvv30}
&=  \sum_{\kappa_1,\ldots,\kappa_N}  e^{-i\sum_j\omega_{\kappa_j}  t} ~ \prod_{j''}\alpha_{\kappa_j''}  ~ \sum_{\kappa'} \hbar\omega_{\kappa'} \hat B^\dag_{\vec\varphi_{\kappa'}}    \hat B_{\vec\varphi_{\kappa'}}\prod_{j'}\hat B^\dag_{\vec\varphi_{\kappa_{j'}} }   |\varnothing\rangle.\phantom{xxxxxxxx}
 \end{align} 
Using the commutation relations 
 $\hat B_{\vec\varphi_{\kappa'}}\hat B^\dag_{\vec\varphi_{\kappa_j}} = \hat B^\dag_{\vec\varphi_{\kappa_j}}\hat B_{\vec\varphi_{\kappa'}} +\delta_{\kappa_j,\kappa'}$ one can show  by recursion 
  that

  \begin{eqnarray} 
 \hat B_{\vec\varphi_{\kappa'}}\prod_{j=1}^N\hat B^\dag_{\vec\varphi_{\kappa_j} }   
      &=& \sum_{j=1}^N \delta_{\kappa',\kappa_j} \prod_{j'\neq j} \hat B^\dag_{\vec\varphi_{\kappa_{j'}} } 
      +  \prod_{j=1}^N\hat B^\dag_{\vec\varphi_{\kappa_j} }   \hat B_{\vec\varphi_{\kappa'}}.
  \end{eqnarray} 
  Indeed, in Eq. \eqref{recursion2} we have shown that it is true for $N=2$. If we assume that it is true for $N$, the following relations show that it is true also for $N+1$:
  \begin{eqnarray} 
 \hat B_{\vec\varphi_{\kappa'}}\prod_{j=1}^{N+1}\hat B^\dag_{\vec\varphi_{\kappa_j} }   
      &=& \left(  \sum_{j=1}^N \delta_{\kappa',\kappa_j} \prod_{j'\neq j} \hat B^\dag_{\vec\varphi_{\kappa_{j'}} } 
      +  \prod_{j=1}^N\hat B^\dag_{\vec\varphi_{\kappa_j} }   \hat B_{\vec\varphi_{\kappa'}}   \right) \hat B^\dag_{\vec\varphi_{\kappa_{N+1}}}  \\
            &=&   \sum_{j=1}^N \delta_{\kappa',\kappa_j} \prod_{j'\neq j} \hat B^\dag_{\vec\varphi_{\kappa_{j'}}}   \hat B^\dag_{\vec\varphi_{\kappa_{N+1}}}
               +  \prod_{j=1}^N\hat B^\dag_{\vec\varphi_{\kappa_j} } \left(   
                        \hat B_{\vec\varphi_{\kappa'}}  \hat B^\dag_{\vec\varphi_{\kappa_{N+1}}}   \right)\\
      &=&   \sum_{j=1}^N \delta_{\kappa',\kappa_j} \prod_{j'\neq j} \hat B^\dag_{\vec\varphi_{\kappa_{j'}}}   \hat B^\dag_{\vec\varphi_{\kappa_{N+1}}}
               +  \prod_{j=1}^N\hat B^\dag_{\vec\varphi_{\kappa_j} } \left(  \delta_{\kappa', \kappa_{N+1}}  
                  +    \hat B^\dag_{\vec\varphi_{\kappa_{N+1}}}  \hat B_{\vec\varphi_{\kappa'}}  \right)\\
      &=&   \sum_{j=1}^N \delta_{\kappa',\kappa_j} \prod_{j'\neq j} \hat B^\dag_{\vec\varphi_{\kappa_{j'}}}   \hat B^\dag_{\vec\varphi_{\kappa_{N+1}}}
      +  \prod_{j=1}^N\hat B^\dag_{\vec\varphi_{\kappa_j} }  \delta_{\kappa', \kappa_{N+1}}  
      +  \prod_{j=1}^{N+1} \hat B^\dag_{\vec\varphi_{\kappa_j}}  \hat B_{\vec\varphi_{\kappa'}}~~~~~~~~~~  \\
      &=&   \sum_{j=1}^{N+1} \delta_{\kappa',\kappa_j} \prod_{j'\neq j} \hat B^\dag_{\vec\varphi_{\kappa_{j'}}}  
      +  \prod_{j=1}^{N+1} \hat B^\dag_{\vec\varphi_{\kappa_j}}  \hat B_{\vec\varphi_{\kappa'}}.
  \end{eqnarray} 
  Thus, using  the fact that  $\hat B_{\vec\varphi_{\kappa'}}|\varnothing\rangle =0$ we have
\begin{eqnarray} 
 \hat B_{\vec\varphi_{\kappa'}}\prod_j\hat B^\dag_{\vec\varphi_{\kappa_j} }   |\varnothing\rangle
      &=& \sum_{j} \delta_{\kappa',\kappa_j} \prod_{j'\neq j} \hat B^\dag_{\vec\varphi_{\kappa_{j'}} }  |\varnothing\rangle 
  \end{eqnarray} 
and further 
\begin{eqnarray} 
 \sum_{\kappa'} \hbar\omega_{\kappa'} \hat B^\dag_{\vec\varphi_{\kappa'}}    \hat B_{\vec\varphi_{\kappa'}}\prod_j\hat B^\dag_{\vec\varphi_{\kappa_j} }   |\varnothing\rangle
   &=& \sum_{\kappa'} \hbar\omega_{\kappa'} \hat B^\dag_{\vec\varphi_{\kappa'}}  \sum_{j}  \delta_{\kappa',\kappa_j} \prod_{j'\neq j}\hat B^\dag_{\vec\varphi_{\kappa_{j'}} }  |\varnothing\rangle \\
   &=& \sum_{j}  \sum_{\kappa'} \hbar\omega_{\kappa'} \hat B^\dag_{\vec\varphi_{\kappa'}}   \delta_{\kappa',\kappa_j}\prod_{j'\neq j} \hat B^\dag_{\vec\varphi_{\kappa_{j'}} }  |\varnothing\rangle  \\
   &=& \sum_{j}  \hbar\omega_{\kappa_j} \hat B^\dag_{\vec\varphi_{\kappa_j}}   \prod_{j'\neq j} \hat B^\dag_{\vec\varphi_{\kappa_{j'}} }  |\varnothing\rangle  \\ \label{bb45}
   &=&\left(  \sum_{j}  \hbar\omega_{\kappa_j} \right) \prod_{{\rm all~}j'} \hat B^\dag_{\vec\varphi_{\kappa_{j'}} }  |\varnothing\rangle.
   \end{eqnarray} 
Finally inserting \eqref{bb45} into \eqref{vvvv30} we obtain
\begin{eqnarray} 
 \nn
\hat H  | \Phi(t)\rangle 
&=&  \sum_{\kappa_1,\ldots,\kappa_N}  e^{-i\sum_j\omega_{\kappa_j}  t} ~ \prod_{j''}\alpha_{\kappa_j''}  ~ \sum_{\kappa'} \hbar\omega_{\kappa'} \hat B^\dag_{\vec\varphi_{\kappa'}}    \hat B_{\vec\varphi_{\kappa'}}\prod_j\hat B^\dag_{\vec\varphi_{\kappa_j} }   |\varnothing\rangle \\
&=&  \sum_{\kappa_1,\ldots,\kappa_N}  e^{-i\sum_j\omega_{\kappa_j}  t} ~ \prod_{j''}\alpha_{\kappa_j''}  
\left(  \sum_{j}  \hbar\omega_{\kappa_j} \right) \prod_{j'} \hat B^\dag_{\vec\varphi_{\kappa_{j'}} }  |\varnothing\rangle \\
    &=&       i\hbar\frac{\partial}{\partial t}  | \Phi(t)\rangle, 
 \end{eqnarray} 
where in the last equality we have used the relation \eqref{dttwo}, which completes the proof.  
\nll
\end{appendices}

{\bf Acknowledgments: }
This work was supported by the ``Investissements d'Avenir'' program, project ISITE-BFC / IQUINS (ANR-15-IDEX-03), QUACO-PRC (ANR-17-CE40-0007-01) and the EUR-EIPHI Graduate School (17-EURE-0002). We also acknowledge support from the European Union's Horizon 2020 research and innovation program under the Marie Sklodowska-Curie grant agreement No. 765075 (LIMQUET).
\nll
\centerline{This article is dedicated to the memory of Bruce W. Shore,} 
\centerline{ who has been a  great  inspiration for us for so many years. }

\vskip1cm

\end{document}